\documentclass[aps,pra,superscriptaddress,showpacs,twocolumn,longbibliography]{revtex4-1}

\usepackage{graphicx}
\usepackage{amsmath}
\usepackage{amsfonts}
\usepackage{xspace}
\usepackage{color}
\usepackage[usenames,dvipsnames]{xcolor}
\usepackage{fixmath}
\usepackage[colorlinks=true,linkcolor=blue,urlcolor=blue,citecolor=blue]{hyperref}
\usepackage{mathtools}
\usepackage[normalem]{ulem}

\DeclareMathOperator{\tr}{\mbox{tr}}

\DeclareMathOperator{\im}{\mbox{Im}}

\newcommand{\se}{Sec.\@\xspace}

\newcommand{\app}{App.\@\xspace}

\newcommand{\ve}[1]{{\bf #1}}

\newcommand{\eq}[1]{Eq.\thinspace{}(\ref{#1})}

\newcommand{\fig}[1]{Fig.\thinspace{}\ref{#1}}

\newcommand{\fc}[1]{({#1})}
\newcommand{\figc}[2]{Fig.\thinspace{}\ref{#1}\thinspace{}\fc{#2}}

\newcommand{\RR}{\ve R}
\newcommand{\CC}{\mathcal{C}}
\newcommand{\dd}{\mathrm{d}}

\newcommand{\Shat}{\hat{\mathbf{S}}}
\newcommand{\shat}{\hat{S}}

\renewcommand{\sp}{\ket{\text{sp}(\ve{Q})}}

\def\ket#1{\mathinner{|{#1}\rangle}}

\newcommand{\rr}{\mathbf{r}}

\newcommand{\qq}{\mathbf{q}}
\newcommand{\kk}{\mathbf{k}}

\newcommand{\QQ}{\mathbf{Q}}

\newcommand{\DD}{\mathcal{D}}

\newcommand{\GG}{\mathcal{G}}

\newcommand{\LL}{\mathcal{L}}

\newcommand{\LLL}{\mathsf{L}}

\newcommand{\GAM}{\mathsf{\Gamma}}

\newcommand{\epss}{\varepsilon}
\newcommand{\phii}{\varphi}
\newcommand{\Phii}{\boldsymbol{\varphi}}

\newcommand{\pitwo}{\pi/2}

\newcommand{\eqfigscl}[2]{\vcenter{\hbox{\includegraphics[scale=#1]{#2}}}}
\newcommand{\eqfigsclbot}[2]{{\hbox{\includegraphics[scale=#1]{#2}}}}

\makeatletter
\newsavebox{\@brx}
\newcommand{\llangle}[1][]{\savebox{\@brx}{\(\m@th{#1\langle}\)}%
  \mathopen{\copy\@brx\kern-0.5\wd\@brx\usebox{\@brx}}}
\newcommand{\rrangle}[1][]{\savebox{\@brx}{\(\m@th{#1\rangle}\)}%
  \mathclose{\copy\@brx\kern-0.5\wd\@brx\usebox{\@brx}}}
\makeatother

\begin{document}

\title{Far-from-equilibrium field theory of many-body quantum spin systems:
Prethermalization and relaxation of spin spiral states in three dimensions}

\author{Mehrtash Babadi}%
\affiliation{Institute for Quantum Information and Matter, Caltech, Pasadena, CA 91125, USA}%
\author{Eugene Demler}%
\affiliation{Department of Physics, Harvard University, Cambridge, MA 02138, USA}%
\author{Michael Knap}%
\affiliation{Department of Physics, Harvard University, Cambridge, MA 02138, USA}%
\affiliation{ITAMP, Harvard-Smithsonian Center for Astrophysics, Cambridge, MA 02138, USA}%
\affiliation{Physik Department, Walter Schottky Institut, and Institute for Advanced Study, Technische Universit\"at M\"unchen, 85748 Garching, Germany}%

\date{\today}

\begin{abstract}

We study theoretically the far-from-equilibrium relaxation dynamics of spin spiral states in the three dimensional isotropic Heisenberg model. The investigated problem serves as an archetype for understanding quantum dynamics of isolated many-body systems in the vicinity of a spontaneously broken continuous symmetry. We present a field-theoretical formalism that systematically improves on mean-field for describing the real-time quantum dynamics of generic spin-$1/2$ systems. This is achieved by mapping spins to Majorana fermions followed by a $1/N$ expansion of the resulting two-particle irreducible (2PI) effective action. 
Our analysis reveals rich fluctuation-induced relaxation dynamics in the unitary evolution of spin spiral states. In particular, we find the sudden appearance of long-lived prethermalized plateaus with diverging lifetimes as the spiral winding is tuned toward the thermodynamically stable ferro- or antiferromagnetic phases. The emerging prethermalized states are characterized by different bosonic modes being thermally populated at different effective temperatures, and by a hierarchical relaxation process reminiscent of glassy systems. Spin-spin correlators found by solving the non-equilibrium Bethe-Salpeter equation provide further insight into the dynamic formation of correlations, the fate of unstable collective modes, and the emergence of fluctuation-dissipation relations. Our predictions can be verified experimentally using recent realizations of spin spiral states with ultracold atoms in a quantum gas microscope~[\href{http://link.aps.org/doi/10.1103/PhysRevLett.113.147205}{S. Hild, \textit{et al.} Phys. Rev. Lett. {\bf 113}, 147205 (2014)}]. 

\end{abstract}


\maketitle

\section{Introduction}

\begin{figure}[th!]
        \centering
        \includegraphics[width=0.45\textwidth]{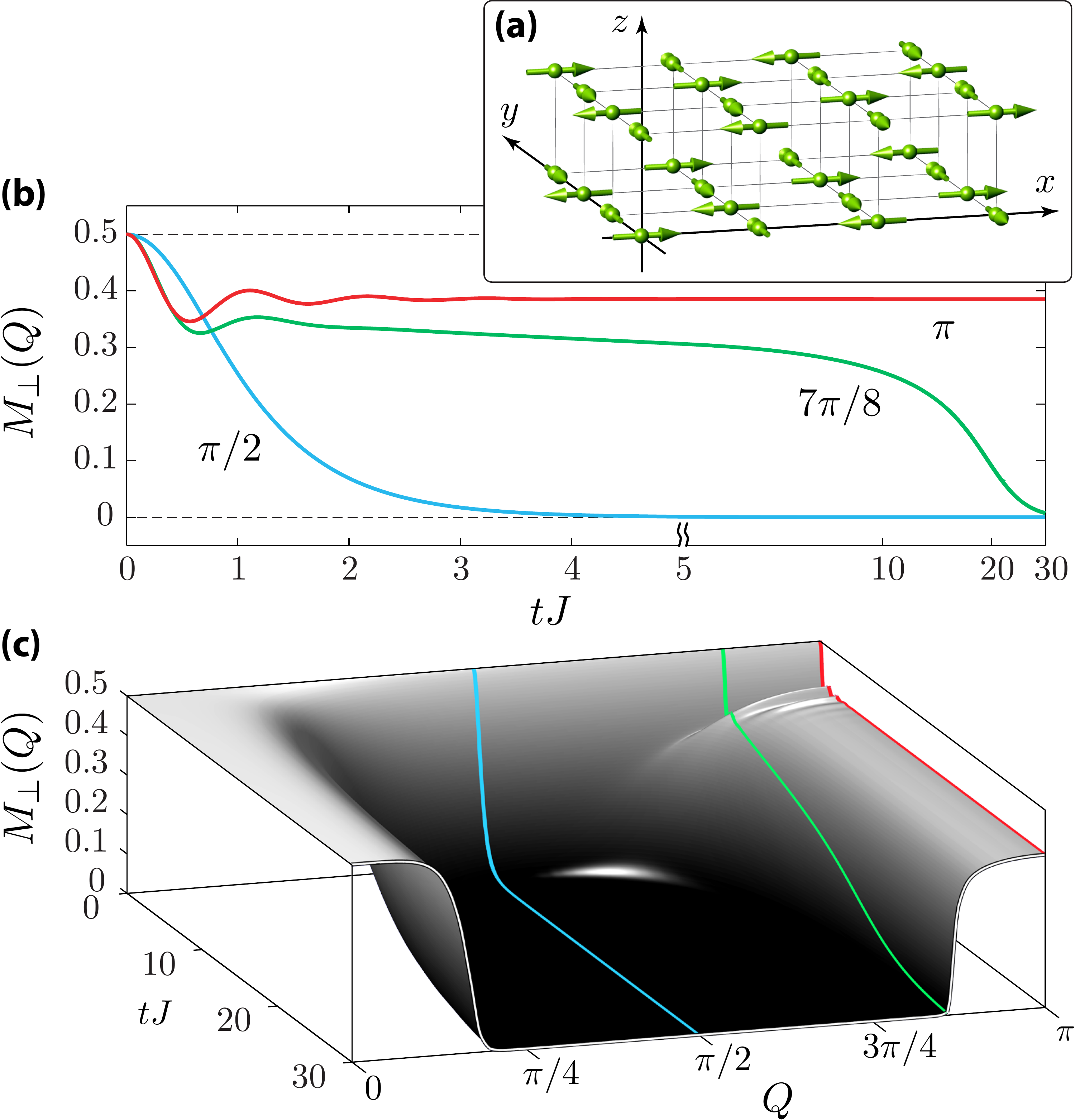}
        \caption{\textbf{Relaxation of spin spiral states in the 3D isotropic Heisenberg model}. \fc{a} The system is prepared in a spin spiral state in the $xy$ plane with the winding $\QQ = (Q, Q, Q)$ as tuning parameter. The figure illustrates the case $Q=\pi/2$. 
        \fc{b} The real-time evolution of the transverse magnetization $M_\perp$ for three different $Q$ as indicated in the plot. For $Q=7\pi/8$, a hierarchical relaxation process emerges with a non-thermal plateau at intermediate times. The time scale is switched to logarithmic at $tJ=5$ for better visibility. \fc{c} A global view of the spiral dynamics. Non-thermal plateaus appear near $Q \sim 0, \pi$.}
        \label{fig:decay} 
\end{figure}

The equilibration of isolated quantum many-body systems is a fundamental and ubiquitous question in physics. It plays a central role in understanding a broad range of phenomena, including the dynamics of the early universe~\cite{traschen1990particle}
, the evolution of neutron stars~\cite{baym1979physics}, pump-probe experiments in condensed matter systems~\cite{goulielmakis_attosecond_2007}, and the operation of semiconductor devices~\cite{haug_quantum_2009}. The simplest perspective on the problem is to recognize a dichotomy between ergodic and non-ergodic systems. The former exhibit fast relaxation to local equilibrium states occurring at microscopic timescales, followed by a slower relaxation process to global thermal equilibrium described by classical hydrodynamics of a few conserved quantities~\cite{hohenberg_theory_1977, mukerjee_statistical_2006, lux_hydrodynamic_2014}. In contrast, non-ergodic systems possess an extensive set 
of conservation laws that prevent their thermalization~\cite{sutherland_beautiful_2004, kinoshita_quantum_2006}.

Recent theoretical and experimental investigations of strongly-correlated systems, however, suggest significant refinements to this dichotomy. For instance, certain systems can be trapped for long times in quasi-stationary ``prethermalized'' states  with properties strikingly different from  true thermal equilibrium~\cite{berges_prethermalization_2004}. Examples include nearly-integrable one-dimensional systems~\cite{gring_relaxation_2012, mitra_correlation_2013, marcuzzi_prethermalization_2013, essler_quench_2014, nessi_glass-like_2015}, and systems with vastly different microscopic energy scales in which slow dynamics results from the slow modes providing configurational disorder and thereby localizing the fast modes~\cite{kagan_1984_localization, schiulaz_ideal_2014, grover_quantum_2014}. Even subtler examples of slow dynamics include the Griffiths phase of interacting disordered systems~\cite{agarwal_anomalous_2014,vosk_theory_2014} and 
translationally invariant systems in higher dimensions with emergent slow degrees of freedom~\cite{berges_prethermalization_2004, moeckel_interaction_2008, berges2008nonthermal, eckstein_thermalization_2009, kollar_generalized_2011, barnett_prethermalization_2011, tsuji2013nonthermal, nessi_2014, smacchia_exploring_2014, bauer_babadi_2014}.\\

In this work, we discuss the emergence of slow dynamics and prethermalization in translationally invariant spin systems that possess continuous symmetries. In higher dimensions, these systems can exhibit thermodynamically stable symmetry broken phases along with gapless Goldstone modes. Here, we show that the relaxation dynamics of low energy initial states that allow symmetry breaking upon thermalization is remarkably different from the relaxation of high energy states, thereby establishing a connection between aspects of equilibrium and non-equilibrium phenomena in these system. In particular, the slow Goldstone modes in the former case result in the emergence of long-lived non-thermal states with a hierarchical relaxation dynamics that closely resembles aging in systems with quenched disorder. A non-perturbative treatment and beyond mean-field corrections are both found to be crucial for describing the relaxation process.

We specifically study the dynamics of the three dimensional (3D) isotropic Heisenberg model initially prepared in a spiral state, see \figc{fig:decay}{a}. Our choice of spin spiral states is motivated by the following considerations. First, the winding of a spiral, $Q$, serves as a tuning parameter for the energy density of the state. The full spectrum of the Heisenberg model is traversed from ferromagnetic (FM) to N\'eel antiferromagnetic (AFM) states upon sweeping $Q$ from $0$ to $\pi$, respectively. In light of the eigenstate thermalization hypothesis (ETH)~\cite{deutsch_quantum_1991, srednicki_chaos_1994, rigol_thermalization_2008}, the energy density of the initial state fully determines the fate of all local observables at late times in ergodic systems. One of the main objectives of this work is to  understand the route toward thermalization of these states. Second, spiral states represent different mean-field solutions of the classical Heisenberg model, all of which are thermodynamically unstable with the exception of $Q=0, \pi$. The fluctuation-induced destruction of the initial order and the emergence of thermodynamically stable ordered or disordered phases at longer times is another question we address here. Lastly, spin spiral states have been recently realized in one and two dimensions using ultracold atoms in a quantum gas microscope~\cite{hild_far--equilibrium_2014,brown_2d_2014}. An extension of these experiments to three dimensions makes a direct experimental scrutiny of our predictions possible.

Our results indicate that spiral states tuned toward FM or AFM states  exhibit a slow hierarchical relaxation and can come arbitrary close to a dynamical arrest, see \figc{fig:decay}{b--c}. Surprisingly, the relaxation dynamics is neither compatible with the trivial relaxation to local thermal equilibrium and slow hydrodynamic evolution, since the spin current is not conserved, nor with the linearized dynamics of the collective modes, which predicts exponentially growing out-of-plane instabilities. In fact, we find the instabilities to self-regulate and slow down significantly. As we elaborate in the following sections, the physical phenomena discussed here are expected to generalize to a broad range of models that exhibit a finite temperature phase transition between a disordered and a symmetry broken phase.

The relaxation of the N\'eel spin spiral state with $Q=\pi$ has been previously studied in the 1D Heisenberg model~\cite{barmettler_relaxation_2009, liu_quench_2014, hild_far--equilibrium_2014, heyl_dynamical_2014}. In contrast to the 3D case studied in the present work, the 1D Heisenberg model does not exhibit a symmetry broken thermal phase and in turn, cannot exhibit the type of prethermalization we discuss here. More recently, the dynamics of the N\'eel state in the Fermi-Hubbard model on an infinite dimensional Bethe lattice has been investigated~\cite{balzer_non-thermal_2015}, however, the approach to the steady state could not be studied due to the small effective exchange interaction.\\

From a technical perspective, our investigation of the non-equilibrium dynamics of spiral states has been enabled by developing a non-perturbative field theoretic formalism applicable to generic spin-$1/2$ systems for arbitrary initial states and geometries, which we refer to as the ``Spin-2PI'' formalism. This is achieved using a Majorana fermion representation of spin-$1/2$ operators~\cite{berezin_1977,tsvelik_new_1992}, enlargement of the spin coordination number by a replica-symmetric extension, and ultimately a systematic $1/N$ fluctuation expansion of the real-time two-particle irreducible (2PI) effective action~\cite{cornwall_effective_1974,berges_introduction_2004}.

The recent rapid progress in the phenomenology of far-from-equilibrium quantum dynamics and its broad applications has been
enabled by similar non-perturbative functional techniques. Examples include 
extensive studies of the $O(N)$ model in non-equilibrium~\cite{berges2002controlled, aarts2002far, cooper2003quantum}, thermalization, prethermalization and non-thermal fixed points~\cite{berges2003thermalization, berges_prethermalization_2004, berges2008nonthermal, schmidt2012non, nowak2014universal}, particle production, reheating and defect generation in inflationary universe models~\cite{berges2003parametric, arrizabalaga2004tachyonic, berges2011quantum, berges2011topological, gasenzer2014gauge}, and dynamics of ultracold fermionic and bosonic gases~\cite{rey2004nonequilibrium, balzer2009nonequilibrium, kronenwett2011far}. The present work is the first to utilize this powerful technique to study the far-from-equilibrium dynamics of interacting quantum spin systems.\\

Understanding the emergence of slow dynamics near thermodynamic phase transitions has implications reaching far beyond the domain of condensed matter physics. For instance, studies of non-equilibrium quantum fields in the context of inflation and early universe dynamics have suggested that the slowing down of quantum evolution near phase transitions is a plausible explanation for the large number of light particles and broken symmetries in the observable universe~\cite{kofman_beauty_2004}.  Given that the experimental verification of theories about early universe phenomena are typically rather indirect, experiments with synthetic many-body systems that allow precise monitoring of real-time dynamics close to phase transitions could play an important role in elucidating the emergence of slow evolution. The dynamics of various interacting spin systems have been already investigated in experiments with synthetic quantum matter, including domain 
formation in spinor condensates~\cite{stenger_spin_1998,sadler_spontaneous_2006}, the precise measurement of the evolution of spin flips in the 
ground state of 1D lattice spin systems~\cite{fukuhara_quantum_2013, fukuhara_microscopic_2013, richerme_non-local_2014, jurcevic_quasiparticle_2014}, quantum coherences in long range models~\cite{yan_observation_2013}, and the relaxation dynamics of spin spiral states in 1D and 2D Heisenberg models~\cite{hild_far--equilibrium_2014,brown_2d_2014}. The experimental observation of the dynamical phenomena discussed here are thus expected to be within close reach.\\

This paper is organized as follows: In \se~\ref{sec:method}, we introduce the Spin-2PI formalism, a technique we develop to study the dynamics of interacting spin systems. Complementary technical details are presented in \app~\ref{sec:supp}. We discuss the relaxation of spin spiral states in \se~\ref{sec:spiral}. The phenomenon of dynamical slowing down and arrest will be presented \se~\ref{sec:arrest}, the long-time thermalization in \se~\ref{sec:thermalization}, and the dynamic formation of correlations and instabilities in \se~\ref{sec:corr}. We conclude our findings in \se~\ref{sec:conclusions}.

\section{The Spin-2PI Formalism\label{sec:method}}
\begin{figure}[t!]
\centering
\includegraphics[width=.48\textwidth]{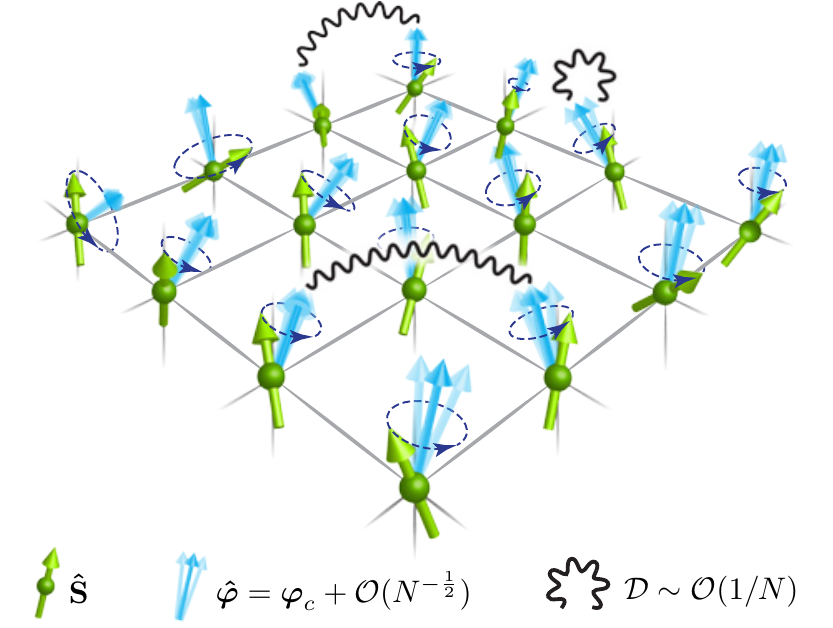}
\caption{\textbf{The Spin-2PI formalism illustrated}. The spins (green arrows) precess about a fluctuating exchange field (uncertain blue arrows). The quantum fluctuations of the exchange field are mediated by a real vector bosons (wiggly lines) and are suppressed by a factor of $1/N$, permitting a systematic expansion.}
\label{fig:formalism} 
\end{figure}   

Consider a generic Hamiltonian describing the pairwise interaction between localized spin degrees of freedom on a given lattice $\LLL$:
\begin{equation}\label{eq:generalH}
\hat{H} = \frac{1}{2}\sum_{j,k \in \LLL} V_{jk}^{\alpha\beta}\, \hat{S}^{\alpha}_j \, \hat{S}^{\beta}_k,
\end{equation}
where $V$ is an arbitrary interaction, $j$ and $k$ denote lattice sites, and $\{\hat{S}^\alpha\}$ are spin-$1/2$ operators.  Summation over the repeated spin indices is assumed. Had $\shat$ been classical angular momentum variables, the Hamiltonian dynamics of the system would be governed by the (non-linear) Bloch equation:
\begin{equation}\label{eq:bloch}
\frac{\dd \mathbf{S}_j}{\dd t} = \boldsymbol{\varphi}_j \times \mathbf{S}_j, \qquad \varphi^{\alpha}_j = \sum_{k \in \LLL} V^{\alpha\beta}_{jk}\,S^{\beta}_k.
\end{equation}
In case of quantum spins, the Bloch equation only describes the evolution of the spin expectation values $\langle \Shat \rangle$ to the extent of which the mean-field Ansatz $\langle \hat{S}_j \hat{S}_k \rangle \approx \langle \hat{S}_j \rangle \, \langle \hat{S}_k \rangle$ is valid. The latter, however, is only justified for lattices with large coordination number, high spin particles, or in the presence of a high-temperature bath. The crucial role of quantum fluctuations in the dynamics of isolated spin-$1/2$ systems in finite dimensional lattices is beyond the reach of semi-classical methods, and demands a more careful treatment.

Here, we propose a formalism for transcending the mean-field approximation for spin evolution by a systematic inclusion of quantum corrections. This is achieved using functional methods and a variant of the large-$N$ expansion technique. As a first step, we construct an auxiliary model in which each spin is replicated $N$ times, and each bond is promoted to $N^2$ bonds between the replicas, with equal weight but with an overall scale factor of $1/N$. The Hamiltonian of the auxiliary model is written as:
\begin{equation}\label{eq:HN}
\hat{H}_{N} = \frac{1}{2}\sum_{j \in \LLL} \left(\sum_{k\in \LLL} V^{\alpha\beta}_{jk}\,\frac{1}{N}\sum_{\sigma'=1}^N \hat{S}_k^{\beta;\sigma'}\right)\,\sum_{\sigma=1}^N\hat{S}_j^{\alpha;\sigma}.
\end{equation}
The initial state $|\Psi_0\rangle$ is also subsequently promoted to an uncorrelated product in the replica space, $\bigotimes_{\sigma=1}^N|\Psi_0\rangle_\sigma$. The original problem is recovered by setting $N=1$. We refer to the sum appearing in the parentheses in Eq.~\eqref{eq:HN} as the {\em exchange field} operator, $\hat{\boldsymbol{\varphi}}_j$, which plays the role of an effective fluctuating magnetic field with which the spins interact. The described large-$N$ construction effectively increases the coordination number of each spin, $z$, to $Nz$, thereby suppressing the fluctuations of $\hat{\boldsymbol{\varphi}}_j$ according to the law of large numbers, $\hat{\boldsymbol{\varphi}}_j = \boldsymbol{\varphi}_{c,j} +\mathcal{O}(1/\sqrt{Nz})$, where $\boldsymbol{\varphi}_{c,j} \equiv \langle \boldsymbol{\varphi}_j \rangle$ is the mean exchange field. In the limit of infinite $N$, the exchange field operator becomes effectively classical such that mean-field dynamics of the original model $\hat{H}$ 
emerges as the asymptotically exact description of 
the dynamics in $\lim_{N\rightarrow \infty} \hat{H}_N$. For large but finite $N$, the fluctuations of $\hat\Phii$ are small but not negligible, and can be systematically incorporated into the dynamics order by order in $1/N$. This program can be carried out within the 
functional method of two-particle irreducible (2PI) effective actions. Crucially, truncating the expansion at a finite order in $1/N$ and taking the limit $N\rightarrow 1$ yields non-perturbative and conserving approximations for the spin dynamics. We refer to this method as the Spin-2PI formalism, which is illustrated schematically in \fig{fig:formalism}. In brief, spins precess about a self-consistently determined exchange mean field, and quantum spin fluctuations are mediated by the local and non-local exchange of a real vector boson whose propagator is suppressed by a factor of $1/N$.

In the remainder of this section, we briefly outline the field theoretical developments that underlie the Spin-2PI formalism. Complementary technical details are given in \app~\ref{sec:supp}. A path integral for the spin-$1/2$ operators is constructed using a representation invoking Majorana fermions~\cite{berezin_1977,tsvelik_new_1992}:
\begin{equation}\label{eq:spinrep}
\hat{\ve S}_j = -\frac{i}{2}\,\boldsymbol{\eta}_j \times \boldsymbol{\eta}_j. 
\end{equation}
The Majorana operators at each site $\{\eta^1_j, \eta^2_j, \eta^3_j\}$ satisfy the Clifford algebra $\{\eta_j^\mu, \eta_k^\nu\} = \delta_{jk} \, \delta^{\mu\nu}$, from which the $SU(2)$ algebra for spins $[\hat{S}^\alpha_j, \hat{S}^\beta_k] = {i}\delta_{jk}\,\epss_{\alpha\beta\gamma} \, \hat{S}^\gamma_j$ and the Casimir condition $\ve{S}_j^2 = {3}/{4}$ follow. The latter ensures a faithful spin-$1/2$ representation without introducing any unphysical states and obviates the necessity of using constraint gauge fields in contrast to the Schwinger slave particle approach~\footnote{For spin systems in thermal equilibrium, the local constraint of Schwinger slave fermions can be removed using a complex chemical potential~\cite{popov1988functional}. This technique is also adapted to the Schwinger-Keldysh formalism in Ref.~\cite{kiselev2000schwinger}.}; see Appendices of Ref.~\cite{coleman1994odd} for a detailed treatment of the Majorana representation for spin-$1/2$ operators.

Replacing the spin operators in $\hat{H}$ using Eq.~\eqref{eq:spinrep}, the Hamiltonian is mapped to that of a many-body system of Majorana fermions with quartic interactions. The large-$N$ program can be identically followed by replicating the slave Majorana particles and assigning a replica index to each. We proceed by constructing a path integral for the Majorana fermions using fermionic coherent states on the closed time path (CTP) Schwinger-Keldysh contour. The Lagrangian is given as:
\begin{multline}
\LL[\eta] = \frac{1}{2}\sum_{j\in \LLL}\sum_{\sigma=1}^N\eta^{\alpha;\sigma}_j\,i\partial_t\,\eta^{\alpha;\sigma}_j + \frac{1}{8N}\sum_{j,k\in\LLL}\sum_{\sigma_1,\sigma_2=1}^N\\
V^{\alpha\beta}_{jk} \, (\boldsymbol{\eta}_j \times \boldsymbol{\eta}_j)^{\alpha;\sigma_1} \, (\boldsymbol{\eta}_k \times \boldsymbol{\eta}_k)^{\beta;\sigma_2}.
\end{multline}
The exchange field is introduced by a Hubbard-Stratonovich decoupling of the quartic term using a real vector boson $\boldsymbol{\varphi}_j$ on each lattice site. The non-equilibrium exchange mean field $\Phii_c$, exchange field fluctuation propagator $\DD$, and the Majorana propagator $\GG$ are introduced as:
\begin{align}
\Phii_c(1) &= \langle \hat{\Phii}(1) \rangle,\nonumber\\
i\DD(1,2) &= \langle T_\CC[\hat{\Phii}(1)\,\hat{\Phii}(2)]\rangle - \Phii_c(1)\,\Phii_c(2),\nonumber\\
i\GG(1,2) &= \langle T_\CC[\eta(1)\,\eta(2)]\rangle.
\end{align}
The integer variables are shorthand for the bundle of lattice site, contour time, spin and replica index. According to Eq.~\eqref{eq:spinrep}, the local spin expectation value is proportional to the fermion tadpole:
\begin{equation}\label{eq:SG}
\langle \hat{S}^\alpha_j(t) \rangle = \frac{1}{2}\,\epss_{\alpha\beta\gamma}\,\GG^{\beta\gamma}_{jj}(t^+,t).
\end{equation}
We obtain the real-time evolution equations for $\GG$, $\DD$ and $\Phii_c$ using the 2PI effective action formalism~\cite{cornwall_effective_1974}. The effective action $\GAM[\GG, \DD, \boldsymbol{\varphi}_c]$ is found by sourcing $\GG$, $\DD$ and $\Phii_c$ and performing Legendre transformations:
\begin{subequations}
\begin{align}\label{eq:gam}
\GAM[\GG, \DD, \Phii_c] &= \frac{1}{2}\tr \ln \GG^{-1} + \frac{1}{2}\tr[\GG_0^{-1}\GG] - \frac{1}{2}\tr\ln \DD^{-1} \nonumber\\&
-\frac{1}{2}\,\tr [\DD_0^{-1} \DD] + \GAM_\mathrm{int}[\GG, \DD, \Phii_c],\\
\label{eq:gam2}
\GAM_\mathrm{int}[\GG, \DD, \Phii_c] &= - \frac{1}{2}\tr[M[\Phii_c] \, \GG] + \frac{i}{2} \Phii_c \DD_0^{-1} \Phii_c+ \GAM_2[\GG,\DD],
\end{align}
\end{subequations}
The bare Majorana and exchange propagators are given as $\GG_0^{-1}(1,2) = i\partial_{t_1}\delta(1,2)$ and $\DD_0^{-1}(1,2) = N\, (V^{-1})\,\delta(t_1,t_2)$, respectively. $M[\Phii_c]$ is the leading order (LO) self-energy (see Eq.~\ref{eq:Mdef}). The evolution equations follow from making $\GAM$ stationary with respect to $\GG$, $\DD$, and $\Phii_c$, see Eqs.~\eqref{eq:SMdys1}-\eqref{eq:SMdys3}.

Save for $\Gamma_2[\GG,\DD]$, the rest of the terms appearing in $\GAM[\Phii_c,\GG,\DD]$ scale as $\mathcal{O}(N)$ and together comprise the leading-order (LO) approximation. The next-to-leading-order (NLO) corrections and beyond are represented by $\GAM_2[\GG,\DD]$ which formally corresponds to the sum of 2PI vacuum diagrams arising from the cubic interaction vertex:
\begin{equation}
\LL_{\rm int}[\eta,\Phii] = \frac{i}{2}\,\Phii \cdot (\boldsymbol{\eta} \times \boldsymbol{\eta})^\sigma = \frac{i}{2}\,\epss_{\alpha\beta\gamma}\,\,\eqfigscl{1.2}{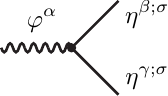}.
\end{equation}
The $1/N$ expansion of $\GAM_{\rm int}$ to the next-to-next-to-leading (NNLO) order is diagrammatically given as:
\begin{equation}\label{eq:Gamint}
\GAM_{\rm int}[\GG,\DD] = \underbrace{\,\eqfigscl{0.75}{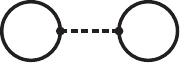}\,}_{\textsf{LO}\,\,\sim\,\,\mathcal{O}(N)} + \underbrace{\,\eqfigscl{0.75}{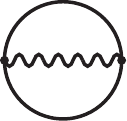}\,}_{\textsf{NLO}\,\,\sim\,\,\mathcal{O}(1)} + \underbrace{\,\eqfigscl{0.75}{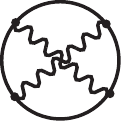} + \eqfigscl{0.75}{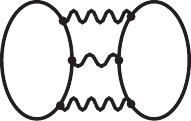} \,}_{\textsf{NNLO}\,\,\sim\,\,\mathcal{O}(1/N)}.
\end{equation}
We have used the stationarity condition, Eq.~\eqref{eq:SMdys3}, to omit $\Phii_c$ in favor of $\GG$ in the LO interaction terms. The Feynman diagram rules are given in Sec.~\ref{app:feyn}.

Truncating the $1/N$ expansion of $\GAM$ at a finite order and setting $N=1$ yields systematic improvements of the mean-field spin dynamics. The ensuing approximate theories are self-consistent and non-perturbative by construction, and respect the conservation laws associated to the global symmetries of the microscopic action, such as magnetization and energy. The latter is crucial for the long-time stability of the non-equilibrium dynamics.

The Bloch equation is recovered upon truncating $\GAM$ at the LO level, see Sec.~\ref{app:details}. Truncations at NLO and beyond give rise to memory effects due to the dynamical fluctuations of the exchange field and result in a two-time Kadanoff-Baym integro-differential equation instead of the mean field Bloch equation, see Eqs.~\eqref{eq:KB1}-\eqref{eq:D2}. Finally, higher order correlators, in particular the spin-spin correlator $i\chi(1,2) \equiv \langle T_\CC [ \hat{S}(1) \hat{S}(2)] \rangle - \langle \hat{S}(1) \rangle \, \langle \hat{S}(2) \rangle$, can be reconstructed with the knowledge of $\GG$ and $\DD$ by solving the non-equilibrium Bethe-Salpeter integral equation on the Schwinger-Keldysh contour, see \app~\ref{app:details}.

We remark that in systems with large spin coordination number $z$, fluctuations of the exchange field are inherently suppressed and the expansion parameter is more accurately identified with $1/(zN)$. Therefore, the large-$N$ expansion of the Spin-2PI effective action in models with $z \gtrsim 1$ is expected to be controlled and rapidly converging, even after taking the limit $N \rightarrow 1$. Studies of the $O(N)$ model show that the most important correction to the mean-field (LO) approximation is captured by the NLO ``fluctuation-exchange'' diagram, along with negligible quantitative corrections from the subleading terms~\cite{aarts2006nonequilibrium,aarts2008effective}.

The replica-based $1/N$ expansion proposed here differs from the usual semi-classical $1/S$ expansion in significant ways even though they improve upon the same mean-field limit. For instance, the replicated Fock space of a single spin is reducible and has many more states compared to a pure spin-$N/2$ representation. A technical advantage of our approach is that it preserves the underlying spin-$1/2$ degrees of freedom, which in conjunction to the Majorana representation leads to the familiar diagrammatic and functional methods. As discussed before, these tools significantly simplify and streamline the calculation of higher order corrections. Furthermore, there is no preferred axis for spin quantization in the Spin-2PI formalism, allowing us to study magnetically ordered and disordered states in a unified way.

\section{Relaxation of spin spiral states in the 3D Heisenberg model\label{sec:spiral}}
In this section, we investigate the unitary evolution of the spin spiral state on a 3D cubic lattice, 
\begin{equation}
 \sp = e^{-i \sum_j\ve{Q}\cdot \RR_j \hat{S}_j^z}\bigotimes_{j\in\mathbb{Z}^3}\ket{\rightarrow}_j,
\end{equation}
under the isotropic Heisenberg Hamiltonian $\hat H=-J \sum_{\langle ij \rangle} \hat{\ve S}_i \cdot \hat{\ve S}_j$ using the Spin-2PI formalism developed in the previous section. Here, $\ket{\rightarrow}_j$ denotes the $x$-polarized state on lattice site $j$. The spiral is prepared in the $xy$-plane with a winding wavevector $\mathbf{Q}$. We assume ferromagnetic couplings $J>0$ for concreteness, even though the sign of the $J$ does not affect the unitary evolution due to the time reversal symmetry of the Heisenberg model.

The spiral state $\sp$ is a simultaneous eigenstate of $\hat{\mathcal{S}}_a(\mathbf{Q}) \equiv \hat{\mathcal{T}}_a \, \hat{\mathcal{R}}_z(Q_a)$, $a=x,y,z$, where $\hat{\mathcal{T}}_a$ and $\hat{\mathcal{R}}_z(Q_a)$ denote the translation by one lattice site along the $a$-axis and rotation by angle $Q_a$ about the $z$-axis, respectively. The translation and rotation symmetries of the isotropic Heisenberg model imply $[\hat{H}, \hat{\mathcal{S}}_a] = 0$, so that the spiral state $\sp$ remains a simultaneous eigenstate of $\hat{\mathcal{S}}_a(\mathbf{Q})$ at all times in the course of unitary evolution. As a result, the out-of-plane magnetization $\langle \hat{S}_j^z(t) \rangle$ vanishes identically, and the spiral magnetic order with the initial winding $\QQ$ persists at all times. The transverse magnetization,
\begin{align}
M_\perp(\QQ,t) &\equiv \frac{1}{L^3} \sum_{j \in \LLL} e^{-i \QQ \cdot \RR_j} \big[\langle \hat{S}_j^x(t) \rangle + i \langle \hat{S}^y_j(t) \rangle\big],
\end{align}
is the only degree of freedom at the level of single spin observables. Also, $M_\perp(\kk,t)=0$ for $\kk \neq \QQ$. We remark that even though the magnetization dynamics is significantly constrained at the level of single spin observables by symmetries, arbitrary spin correlations are allowed to form in the course of evolution, including both in- and out-of-plane spin correlations at arbitrary wavevectors.

A simplifying aspect of the present problem is that the apparently broken translation symmetry of the spiral state can be restored using an ``unwinding'' unitary transformation $\hat{U}_\QQ \equiv e^{i \sum_j\ve{Q}\cdot \RR_j \hat{S}_j^z}$ under which the spiral state transforms into a uniform $x$-polarized product state $|\tilde{\Psi}_0\rangle = \hat{U}_\QQ\sp = \bigotimes_j\ket{\rightarrow}_j$. The unwinding transformation, however, transforms the Hamiltonian $\hat{H} \rightarrow \tilde{H} = \hat{U}_\QQ \hat H \hat{U}_\QQ^\dag$ to an anisotropic Heisenberg model with a Dzyaloshinskii-Moriya term:
\begin{multline}\label{eq:Hspiral}
\tilde{H} = - J\sum_{\langle j, k \rangle}\Big[\hat{S}^z_j \hat{S}^z_k + \cos \mathbf{Q}\cdot(\RR_j-\RR_k)\,\big(\hat{S}^x_j \hat{S}^x_k + \hat{S}^y_j\hat{S}^y_k\big)\\
- \sin \mathbf{Q}\cdot(\RR_j-\RR_k)\big(\hat{S}^x_j \hat{S}^y_k - \hat{S}^y_j \hat{S}^x_k\big)\Big].
\end{multline}
The translation invariance of the initial state in the spiral frame significantly simplifies the structure of the Spin-2PI equations: $\GG$ and $\Sigma$ become local in the real space while $\DD$ depends only on the distance between the sites. These simplifications hold for arbitrary truncations of $\GAM_\text{int}$. Additionally, the bosonic self-energy $\Pi$ becomes local in the real space at the NLO truncation. The magnetization is non-vanishing only along the $x$-direction in the spiral frame due to the symmetry considerations mentioned earlier. The quantities calculated in the spiral frame can be readily transformed to the lab frame using appropriate rotations. In particular, \eq{eq:SG} gives $M_\perp(\QQ,t) = (1/2)\,\GG^{23,>}(t,t)$ with $\GG$ calculated in the spiral frame. We choose the winding to be along the diagonal direction $\mathbf{
Q} = (Q,
Q,Q)$ hereafter and refer to the spiral winding with the single scalar $Q \in [0,\pi]$.\\

At the LO level, the spin dynamics is governed by the Bloch equation, \eq{eq:bloch}. The exchange mean field $\Phii_c$ is parallel to the local magnetization at all lattice sites in a spiral state, implying the absence of any dynamics. In other words, the spiral states are fixed points of the mean-field dynamics for all windings $Q$.

Going beyond the LO dynamics and including the exchange field fluctuations by taking into account the NLO corrections, the spiral state exhibits an intriguing fluctuation-induced relaxation dynamics. States with different windings have different energy densities, along with different strength of in-plane and out-of-plane spin fluctuations, and are found to relax in strikingly different ways. As we discuss below, these factors conspire to give rise to a non-trivial hierarchical relaxation scenario for spiral states lying close to thermodynamically stable orders, exhibiting prethermalization~\cite{berges_prethermalization_2004}, and 
dynamical arrest resembling glassy systems~\cite{Cugliandolo_slow_2003}.

\subsection{Relaxation and dynamical arrest of the transverse magnetization\label{sec:arrest}}

The spiral state for $Q=0$ is a fully polarized FM eigenstate of the Heisenberg model and is therefore stationary. The $Q = \pi$ spiral, on the other hand, corresponds to an uncorrelated N\'eel state which in three dimensions has a large overlap with the correlated AFM state lying at the upper end of the spectrum of the FM Heisenberg model. As a result, the system is expected to achieve a steady state marked with a finite staggered magnetization after a short course of dephasing dynamics, provided that the generated effective temperature is below the ordering temperature.
The evolution of $M_\perp$ is shown in \figc{fig:decay}{b} for several choices of $Q$, along with a global surface plot for $Q \in [0, \pi]$ and $tJ \in [0, 30]$ in~\figc{fig:decay}{c}. The stationarity of the FM state ($Q=0$) and the rapid settlement of N\'eel state ($Q=\pi$) to a steady state with finite staggered magnetization is observed.\\

\noindent\textit{Short-time dephasing dynamics---} For all $Q$, the first stage of dynamics is a short-time relaxation of the form $M_\perp \approx 1/2 - \nu_Q t^2$ arising from the dephasing between the eigenstates that overlap with the spiral. A straightforward calculation using the short-time expansion $\langle \Shat(t) \rangle = \langle \Shat \rangle_0 + it \, \langle[\hat{H},\Shat]\rangle_0 + \frac{(it)^2}{2}\big\langle \big[\hat{H},[\hat{H},\Shat]\big] \big\rangle_{0} + \ldots$ gives $\nu_Q = \frac{3}{8} \, J^2 \, (\cos Q-1)^2$. The values of $\nu_Q$ extracted from the numerically obtained $M_\perp$ are in agreement with the exact result, see Fig.~\ref{fig:comp}. The second stage of relaxation dynamics depends on the winding of spiral and is either directly thermalizing, or exhibits long-lived prethermalized states preceding the true thermalization. We discuss these cases separately.\\

\noindent\textit{Spiral states with $Q \sim \pi/2$---} Spin spiral states with $Q \sim \pi/2$ have a high energy density with respect to both the FM and the AFM state. Thus, $Q \sim \pi/2$ spiral states overlap with a large number of eigenstates of the Heisenberg model. Such a broad superposition of states lead to fast dephasing which is found to be within a few exchange times. Our results indicate a rapid onset of exponential decay $M_\perp \sim e^{-\gamma_Q t}$ with the fastest rate occurring at $Q = 0.55(1) \pi \sim \pi/2$.\\

\noindent\textit{Spiral states with $Q \sim 0$ and $Q \sim \pi$---} A complex multi-scale relaxation scenario emerges for spirals with windings tuned to $Q \sim 0$ and $Q \sim \pi$, lying close to FM and AFM magnetic orders, respectively. The transverse magnetization exhibits an intermediate plateau for these initial states which appears continuously upon tuning $Q$, see \fig{fig:decay}. The plot of $M_\perp$ shown in \figc{fig:decay}{b} for $Q=7\pi/8$ displays the intermediate plateau followed by relaxation at later times. As $Q$ is tuned closer toward $0$ or $\pi$, the lifetime of the plateau increases abruptly and the magnetization comes to a dynamical arrest. We investigate the nature of such long-lived plateaus in more detail in the following sections.

\subsection{Prethermalization vs. Thermalization\label{sec:thermalization}}
\begin{figure}
        \centering
        \includegraphics[width=0.48\textwidth]{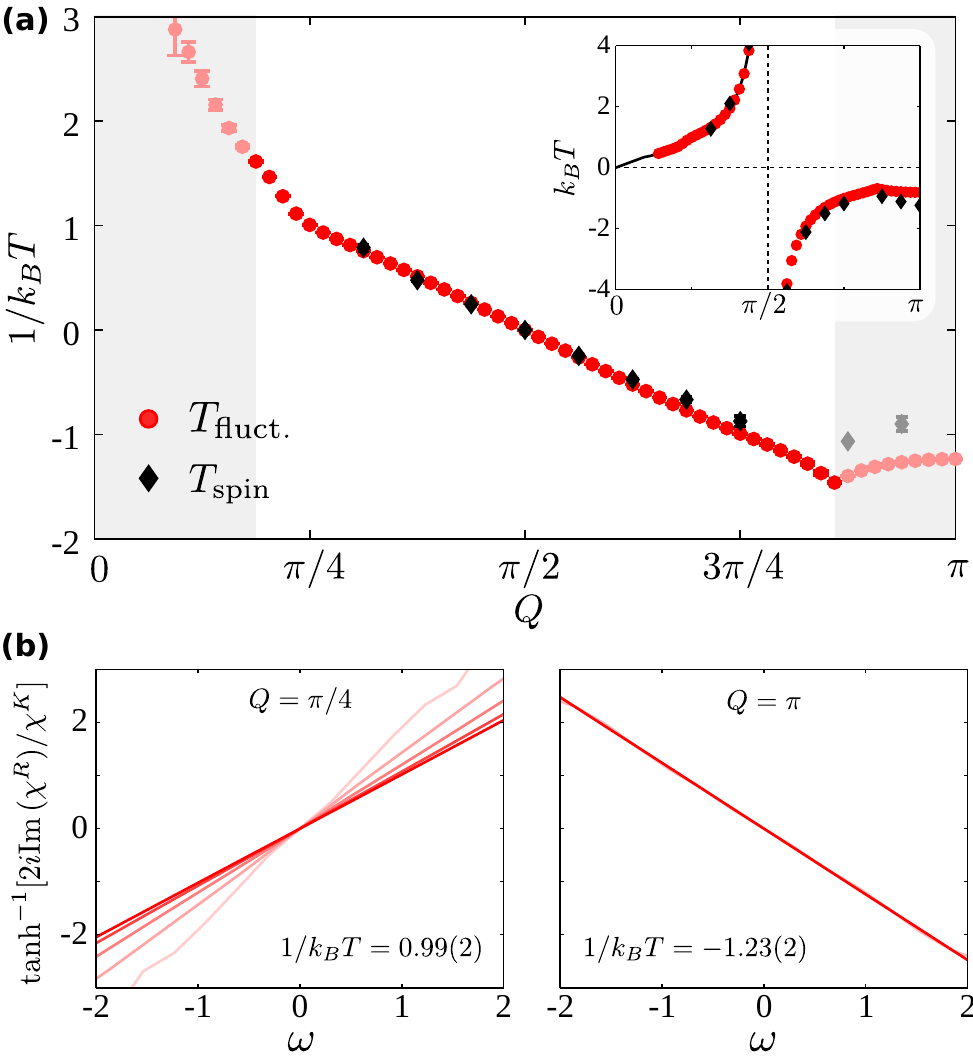}
        \caption{ \textbf{Thermalization of the spin spiral state.}  \fc{a} The effective inverse temperature of local spin $T_\text{spin}$ and local exchange field fluctuations $T_\text{fluct.}$ obtained from the fluctuation-dissipation relations in steady state. The two temperatures are in agreement for spiral windings near $Q \sim \pi/2$, supporting the true thermalization of the system. The temperatures calculated in the prethermalized plateaus $Q\sim 0,\pi$ (shaded regions) disagree with each other, and generically differ from the temperature of the true thermal states that emerge at later times. 
        Inset: the temperature $k_BT$ as a function of $Q$ (same data as in the main panel) displays a resonance from positive infinite temperature to negative infinite temperature at the classical duality point $Q=\pi/2$. \fc{b} The approach of $T_\text{fluct.}$ to steady state (light to dark) as 
obtained from fluctuation-dissipation relations for $Q=\pi/4$ (left) and $Q=\pi$ (right). The steady state temperatures are shown on the plots.
        }
        \label{fig:Teff} 
\end{figure}   
\begin{figure*}[t]
        \centering
        \includegraphics[width=0.98\textwidth]{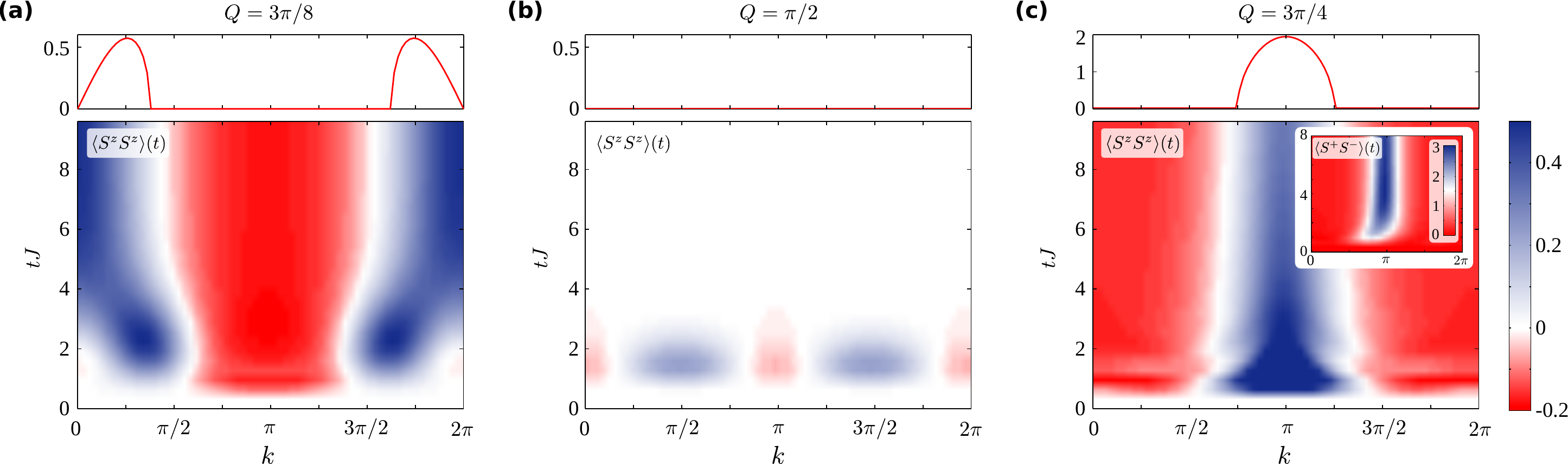}
        \caption{\textbf{The evolution of spin correlations}. Top panels: Growth rate of out-of-plane instable modes obtained from a linear response analysis. Bottom panels: Numerically calculated correlation function $\langle \hat S_\kk^z \hat S_{-\kk}^z \, \rangle(t)= i\chi^{zz,K}_\kk(t,t)$ as a function of the lattice wavevector $\kk=(k,k,k)$ within the Spin-2PI formalism including NLO corrections. \fc{a} $Q=3\pi/8$, \fc{b} $Q=\pi/2$, and \fc{c} $Q=3\pi/4$.  The inset in \fc{c} shows the connected part of the in-plane correlations $\langle \hat S^+_\kk \hat S^-_{-\kk} \rangle(t)$. }
        \label{fig:corr} 
\end{figure*}

Due to the non-integrability of the 3D Heisenberg model, the energy distribution of spin fluctuations is expected to approach a thermal population in the long time limit, according to the eigenstate thermalization hypothesis (ETH)~\cite{deutsch_quantum_1991, srednicki_chaos_1994, rigol_thermalization_2008}. We investigate the nature of steady states emerging in the dynamics by calculating the spin-spin correlation and response functions, corresponding to the Keldysh (K) and retarded (R) components of the CTP spin-spin correlator $\chi(t,t')$, by solving the non-equilibrium Bethe-Salpeter equation (see \app~\ref{app:details}). At thermal equilibrium, these quantities are related via the bosonic fluctuation-dissipation relation (FDR):
\begin{equation}
 i\chi^K(\omega)=-2 \coth(\omega/2 k_B T) \im[\chi^R(\omega)]\;, 
 \label{eq:fdr}
\end{equation}
where $T$ is the effective temperature. Here, $\omega$ refers to the Fourier frequency in the time difference $t-t'$ in the steady state achieved at long times. Likewise, one can define an effective temperature for the exchange field fluctuations using the bosonic FDR between $\DD^K$ and $\DD^R$. We refer to the temperatures obtained from local spin $\chi$ and exchange fluctuations $\DD$ as $T_\mathrm{spin}$ and $T_\mathrm{fluct.}$, respectively.

The effective temperatures obtained from the FDR in the steady state are shown in \figc{fig:Teff}{a}. For all spiral windings $Q$, we find that FDR is satisfied to an excellent degree for both local spin and exchange fluctuation correlators once the steady state is reached, see \figc{fig:Teff}{b}. 
However, as we discuss below, the effective temperature obtained from spin and exchange fluctuations may disagree with each other. This allows us to distinguish prethermalization from true thermalization.\\

\noindent\textit{Thermalization of spiral states with $Q \sim \pi/2$---} For a range of spiral wavevectors $\pi/4 \lesssim Q\lesssim 3\pi/4$, the steady state temperatures obtained from all bosonic modes, i.e. local and non-local in- and out-of-plane spin and exchange field fluctuations, agree with each other, suggesting the complete thermalization of the system and in accordance with the ETH.

Spirals with $Q=\pi/2$ flow to an infinite temperature thermal state, which is understood from the duality $Q \to \pi - Q$, $J \to -J$ present in the classical Heisenberg model. This classical duality extends to the quantum Heisenberg model in the high temperature regime. The duality point $Q=\pi/2$ further marks the resonance from positive temperatures for $Q<\pitwo$ to negative temperatures for $Q>\pitwo$, see the inset of \figc{fig:Teff}{a}. The $T<0$ thermal states of the FM Heisenberg model with coupling $-|J|$ corresponds to $T>0$ states of the AFM Heisenberg model with coupling $+|J|$, and vice versa. Negative temperature states naturally arise in isolated systems with bounded energy spectra as legitimate thermal states and occur when the initial energy density lies closer to the upper edge of the energy spectrum. \\

\noindent\textit{Prethermalization of spiral states with $Q \sim 0,\pi$---} For spiral states with $Q \sim 0,\pi$, where the system develops a prethermal plateau, the effective spin and exchange field fluctuation temperatures disagree, even though the FDR is satisfied well for each mode individually. This finding supports the prethermalized nature of such steady states. It is understood that the temperatures calculated within the prethermal plateau [shown as shaded regions in \figc{fig:Teff}{b}] correspond to the effective temperature of individual modes, and not the true thermodynamical temperature. We expect the two temperatures to approach each other at longer times once the system exits the prethermalized plateau and progresses toward a fully thermalized state.

The spiral state with $Q=0$ is an exact ground state of the system and FDR yields $T=0$ as expected. In contrast, the $Q=\pi$ state approaches a finite temperature, which is understood by the fact that the uncorrelated N\'eel state must be ``dressed'' with spin correlations before the steady state is reached, see inset of \figc{fig:Teff}{a}. The disparity between the evolution of $Q=0$ and $Q=\pi$ states reveals the quantum mechanical nature of spins, and the breakdown of the classical duality $Q \rightarrow \pi - Q$ in the low temperature regime.\\

The 3D Heisenberg model exhibits a finite temperature equilibrium phase transition from the disordered paramagnetic phase to the ordered FM or AFM phase, depending on the sign of the exchange coupling $J$. For the spiral at $Q=\pi$, the FDR of the spin fluctuations are not well fulfilled at accessible times while those for exchange field fluctuations are. The temperature extracted from the latter $|T_{\rm fluct.}(Q=\pi)| = 0.82J$ lies below the the AFM ordering temperature $T_c^\text{AFM}=0.946(1)J$. The latter has been obtained from quantum Monte Carlo simulations~\cite{sandvik_critical_1998}.

Crucially, the near-thermal distribution of fluctuations in the prethermalized plateaus, the stability of FM/AFM ordered phases at finite energy densities in the 3D Heisenberg model, and the proximity of $Q \sim 0, \pi$ spiral states to these stable orders allow us to draw a connection between the long-time stability of such spiral states and spontaneous symmetry breaking at equilibrium: the spiral winding $Q$ sets the energy density of the system and subsequently the effective temperature $T(Q)$ in the prethermal state. $T(Q)$ approximately dictates the magnitude of spin fluctuations on the top of the spiral states which locally resemble either FM or AFM for $Q \sim 0, \pi$. Depending on $Q$, $T(Q)$ can either lie below or above the critical transition temperature, $T_c^\text{FM}$ or $T_c^\text{AFM}$, thereby providing an approximate condition for the local stability of the spiral order. We will study the global instability of the spiral states and their destruction at longer times in the next section.

According to the above discussions, the connection made between emergence of slow dynamics and symmetry breaking at equilibrium essentially hinges on the eigenstate thermalization hypothesis and the Mermin-Wagner theorem. Therefore, this connection is expected to reach beyond the present discussion, and to generalize to a broader range of initial states and models that exhibit spontaneous continuous symmetry breaking.

\subsection{Instabilities and Correlations\label{sec:corr}}
\begin{figure*}[th!]
\centering
\includegraphics[width=0.98\textwidth]{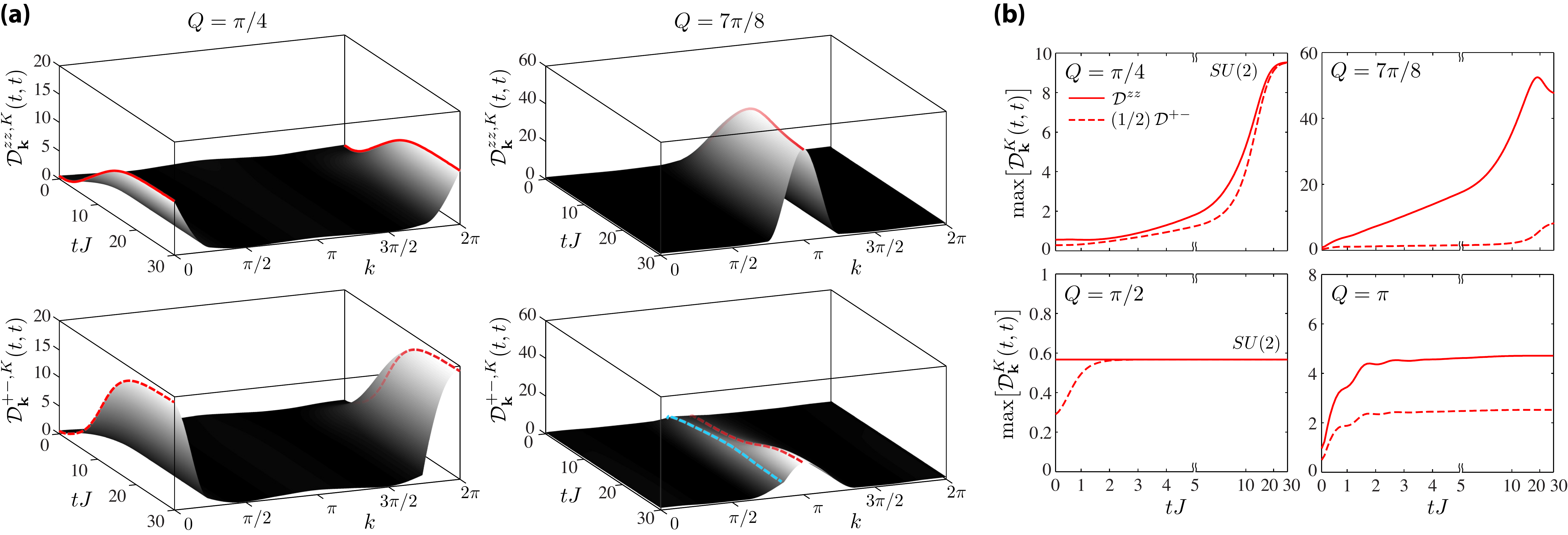}
\caption{\textbf{Dynamics of exchange field fluctuations.} \fc{a} The out-of-plane (top) and in-plane (bottom) exchange field fluctuations as a function of time and momentum $\mathbf{k}=(k,k,k)$ for $Q=\pi/4$ (left) and $Q=7\pi/8$ (right). The red lines indicate the most enhanced mode in the long time limit; the blue dashed line in the lower right plot corresponds to the $k=Q$ in-plane mode which initially exhibits the strongest enhancement of correlations. \fc{b} The evolution of the late-time most enhanced mode for $Q=\pi/4, \pi/2$ (left) and  $Q = 7\pi/8, \pi$ (right). In cases where the system thermalizes, left column, $SU(2)$ symmetry emerges in the long time limit, while it is broken in the prethermal case $Q=7\pi/8$ and for $Q=\pi$, right column. In the latter case, the system can exhibit true long-range order provided its effective temperature is below the critical temperature of the equilibrium phase transition and thus be thermal and simultaneously break $SU(2)$ symmetry.}
\label{fig:D} 
\end{figure*}

The dynamical stabilization of spirals near the FM and AFM orders, and consequently the appearance of prethermal plateaus, was understood on the basis of thermodynamical arguments in the previous section. However, even though the spiral states are fixed points of the mean-field dynamical equations, they are unstable and have a tendency to form out-of-plane textures as the energy of spiral states can be reduced by an appropriate out-of-plane tilt. Therefore, in a thermodynamical ensemble where arbitrary out-of-plane fluctuations are allowed, these saddle points fail to give rise to symmetry broken states, leaving $Q=0$ and $Q=\pi$ as the only thermodynamically stable orders in the Heisenberg model. Therefore, the present situation must be regarded from the perspective of quantum dynamics, i.e. the unitary evolution of a pure state $\sp$ rather than the statistical fluctuations in a mixed thermodynamical ensemble. Here, the system remains in a pure state at all times and the magnetic order is confined to the 
$xy$ spiral plane due to the symmetries discussed at the beginning of Sec.~\ref{sec:spiral}. It is therefore conceivable that symmetry-protected dynamical constraints allow thermodynamically unstable saddle points to become long-lived states in the course of unitary dynamics.\\

\noindent\textit{Out-of-plane instability.---} The out-of-plane instability of the spiral state in the Heisenberg model can be studied either by performing a linear response analysis of the Bloch equations, or similarly from the Holstein-Primakoff spin-wave analysis. Either way, the dispersion of out-of-plane spin-waves forming on the top of the spiral is found as $\omega_\kk=\sqrt{\epsilon_\kk^2 -\Delta_\kk^2}$~\cite{hild_far--equilibrium_2014,conduit2010dynamical}, where:
\begin{allowdisplaybreaks}
\begin{align}
\epsilon_\kk &= -JS \sum_{d=1}^3 \left[(1+\cos \QQ \cdot \hat{\mathbf{e}}_d) \, \cos \kk \cdot \hat{\mathbf{e}}_d - 2 \cos \QQ \cdot \hat{\mathbf{e}}_d\right],\nonumber\\
\Delta_\kk &= -JS \sum_{d=1}^3 \left[(1-\cos \QQ \cdot \hat{\mathbf{e}}_d)\cos \kk \cdot \hat{\mathbf{e}}_d\right].
\end{align}
\end{allowdisplaybreaks}
Unstable modes arise when $\omega_\kk$ assumes imaginary values. Except for $Q=0,{\pi}/{2},\pi$, one always finds such unstable modes: for $Q<{\pi}/{2}$, the fastest growing mode is $\mathbf{k}=(k,k,k)$ with $k=\cos^{-1}[\cos^2(Q/2)]$ along with a sharp cutoff $|k| \leq Q$; for $Q>{\pi}/{2}$, unstable modes occur for $|k-\pi| \leq Q$, with the fastest mode always being the staggered $k=\pi$ mode, independently of $Q$.

A simple estimate for the lifetime of the prethermal plateaus is obtained by calculating the time it takes for typical unstable out-of-plane collective mode to grow to $\mathcal{O}(1)$. The rationale behind this estimate is that the in-plane order can not coexist with strong enough out-of-plane fluctuations. Expanding around $Q=0,\pi$, we obtain a scaling $t \sim 1/Q^2$ for the FM-like and $t \sim 1/(\pi-Q)$ for the AFM-like spirals, up to logarithmic corrections. However, the lifetime of plateaus as found from the Spin-2PI formalism exceeds the above estimates; in particular, as $Q$ is tuned closer toward $0$ or $\pi$, we observe a faster increase of the lifetime of prethermal plateaus. As we discuss below, the increased lifetime can be explained on the basis of the self-regulation of out-of-plane spin fluctuations.\\

The top panels in \fig{fig:corr} show $\mathrm{Im}[\omega_\kk]$ for several values of $Q$, along with the evolution of equal-time out-of-plane spin correlations $i\chi_\kk^{zz,K}(t,t) = \langle \hat{S}^z_{\kk}(t)\,\hat{S}^z_{-\kk}(t)\rangle$ obtained by solving the non-equilibrium Bethe-Salpeter equation in the NLO approximation, bottom panels. Further, the connected part of in-plane correlations $i\chi_\kk^{+-,K}(t,t) = \langle \hat{S}^+_{\kk}(t)\,\hat{S}^-_{-\kk}(t)\rangle - \langle \hat{S}^+_{\kk}(t) \rangle \langle \hat{S}^-_{-\kk}(t) \rangle $ is shown in the inset of panel \fc{c}.

At $t=0$, spin correlations are zero in accordance with the initial spiral state $\sp$ being an uncorrelated product state. The out-of-plane correlations form at times $t \sim J$. The most enhanced correlations coincide with the wavevector predicted by the linear response analysis to a good degree. The sharp cutoffs predicted by this analysis are found to be smeared, which is expected due to the mode coupling embedded in our 
self-consistent approach. The time scale for the formation of correlations is found to be on the order of the dephasing time, reflecting the fact that the short-time dephasing dynamics and formation of correlations are manifestations of the same phenomenon.

For spiral states that thermalize within the numerically achievable time scales, we observe a smooth shift in both in-plane and out-of-plane spin correlations from the initial $Q$-dependent enhanced modes to either $k=0$ or $k=\pi$, depending on whether $Q<{\pi}/{2}$ or $Q>{\pi}/{2}$, respectively [see~\fig{fig:corr}(a), (c), and the inset]. Even though the linear response analysis correctly indicates the wavevector of the fastest growing out-of-plane mode, the spin correlations rapidly saturate to their maximum values, as opposed to an unbounded exponential growth. 
A similar rapid dynamical regulation of the growth of unstable modes was previously reported in Ref.~\cite{berges2003parametric,berges2008nonthermal} in the context of parametric resonance in the $O(N)$ model. In the present context, this phenomenon explains why the lifetime of the plateaus exceeds the estimate obtained from the linear response analysis, and indicates the important role of mode coupling between spin waves and the necessity of non-perturbative treatments.

For spiral states that exhibit long-lived prethermal plateaus, we study the exchange field correlations $\mathcal{D}$, a quantity that is closely related to $\chi$ but can be calculated for much longer times with less computational resources. The evolution of $\mathcal{D}^{zz,K}_\kk(t,t)$ and $\mathcal{D}^{+-,K}_\kk(t,t)$ for $Q=\pi/4$ and $Q=7\pi/8$ are shown in \figc{fig:D}{a}. The former corresponds to a spiral state that thermalizes at about $20J^{-1}$, while the latter exhibits a magnetization plateau up to $\tau_M \sim 20 \, J^{-1}$, as shown in~\fig{fig:decay}. For $t \lesssim \tau_M$, the most enhanced in-plane mode occurs at $k=Q$ which upon demagnetization smoothly switches to $k=\pi$ for $t \gtrsim \tau_M$. The most unstable out-of-plane mode is always at $k=\pi$. 

As a further check for thermalizing behavior, we study the restoration of the $SU(2)$ symmetry in the exchange field fluctuations $\DD^{zz,K}_\kk(t,t) \rightarrow (1/2)\DD^{+-,K}_\kk(t,t)$, see \figc{fig:D}{b}. For $Q=\pi/4$ and $Q=\pi/2$, we find that the $SU(2)$ symmetry is restored at longer times (left column), while for $Q=7\pi/8$ and the N\'eel initial state $Q=\pi$, the $SU(2)$ symmetry remains broken at all accessible times (right column). Notably, the out-of-plane fluctuations for $Q=7\pi/8$ is found to be an order of magnitude stronger than the $Q=\pi$, in agreement with the previously mentioned existence of an unstable out-of-plane mode for the former state and its absence in the latter.

The magnitude of in-plane fluctuations remain essentially constant in the plateau for $Q=7\pi/8$ (top right) while the out-of-plane fluctuations monotonically increase and reach a maximum at $t \sim 20\,J^{-1} \sim \tau_M$, precisely when the prethermal magnetization decays. This finding connects the decay of the the spiral to the growth of out-of-plane fluctuations. The time $\tau_M$ also marks a reversal in the trend of out-of-plane and in-plane correlations. Even though this change indicates a first step toward establishing $SU(2)$-symmetric correlations, the condition is far from being satisfied at $t \sim \tau_M$ and is bound to occur on much longer time scales, indicating a hierarchical relaxation scenario with the relaxation of magnetization preceding the relaxation of correlations. 

The appearance of long-lived prethermal states and the hierarchical relaxation is reminiscent of aging dynamics in classical structural glass models with quenched disorder~\cite{Cugliandolo_slow_2003} and kinematically constrained models~\cite{ritort_glassy_2003}. Similar multi-scale glassy relaxation dynamics has been recently reported in the quench dynamics of fermions in a nearly-integrable 1D model using a different method~\cite{nessi_glass-like_2015}.\\

\begin{figure}[t!]
\centering
\includegraphics[width=0.48\textwidth]{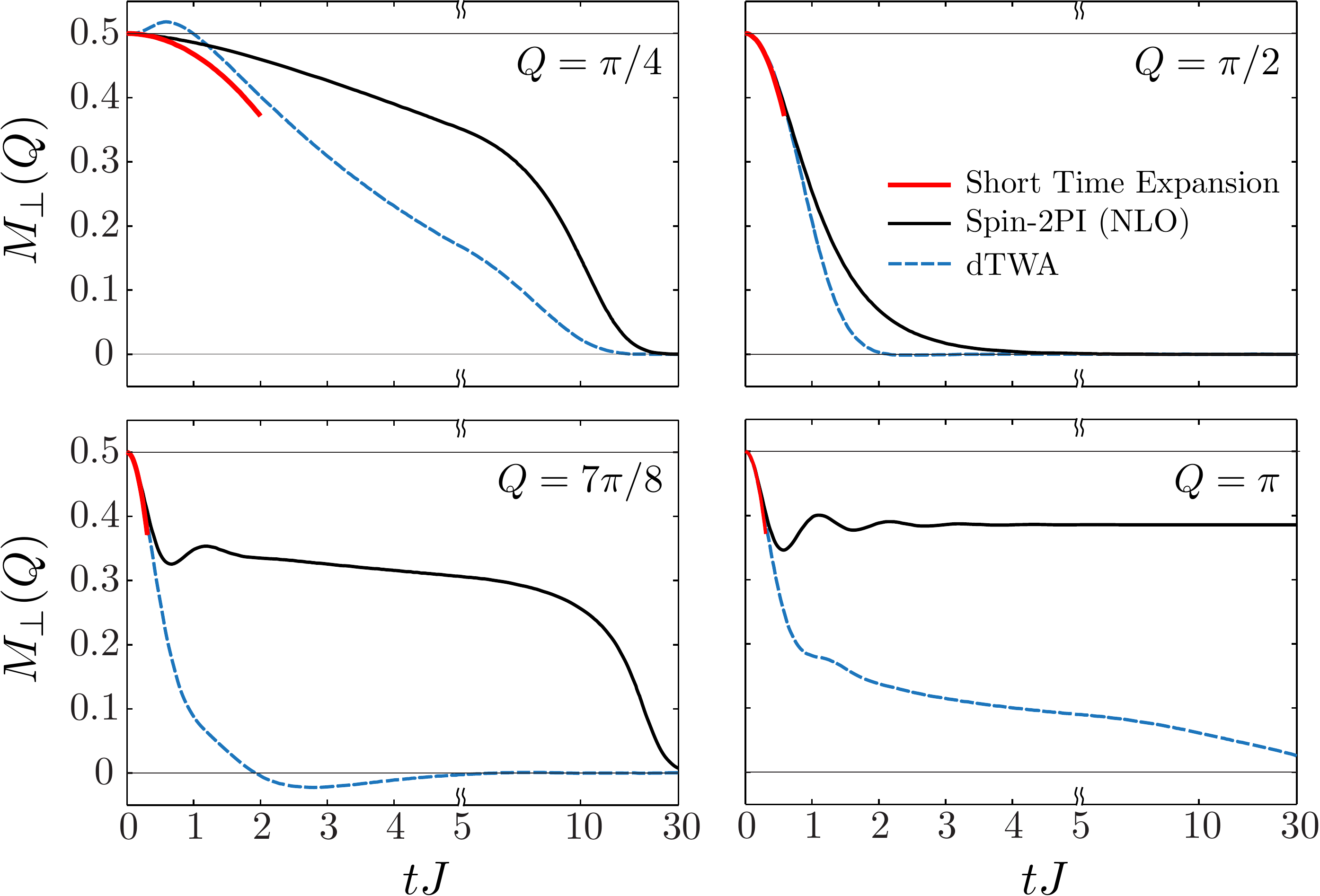}
\caption{\textbf{Comparison between Spin-2PI and semi-classical dynamics}. The Spin-2PI results (solid black lines) are compared to the semi-classical dynamics obtained from the dTWA (dashed blue lines). The short time analytic result from Sec.~\ref{sec:arrest} is also shown for reference (thick red lines). The long time dynamics of the two methods are significantly different. In particular, the dTWA is not capable of describing the long-lived prethermal plateaus in contrast to the Spin-2PI formalism. The scale of the time axis is switched from linear to logarithmic at $tJ = 5$ for better visibility.}
\label{fig:comp} 
\end{figure}

\noindent\textit{Comparison to semi-classical methods---} According to the discussions presented so far, the relaxation dynamics of the spiral state accompanies the formation of in- and out-of-plane quantum correlations in the system. In order to study the role of correlations further, we compare our predictions with the results obtained from the discrete truncated Wigner approximation (dTWA)~\cite{wootters_wigner-function_1987,schachenmayer_many-body_2015}, a variant of the semi-classical TWA method~\cite{polkovnikov_phase_2010} that relies on mean-field trajectories. The magnetization obtained from Spin-2PI (solid black lines) and dTWA (dashed blue lines) are compared in \fig{fig:comp}. The two methods generically agree with the analytic short time expansion (red thick lines), with the exception that dTWA does not reproduce the correct short time dynamics for small $Q$, \textit{cf.} $Q=\pi/4$ in \fc{a}~\footnote{Different trajectory sampling schemes yields slightly different results. Here, we use the correlated sampling scheme described in Ref.~\cite{wootters1987wigner}. An uncorrelated sampling alleviates the short time unphysical behavior. However, with neither scheme dTWA exhibits the plateau behavior.}. The two methods, however, predict strikingly different long time dynamics. Even though dTWA exhibits some degree of dynamical slowing down for FM-like 
and AFM-like spirals, it produces neither the prethermal plateau for $Q=7\pi/8$, nor the finite steady-state magnetization for $Q=\pi$. We note that the latter is supported by exact QMC calculations.

\section{Conclusions and Outlook\label{sec:conclusions}}
We formulated a non-perturbative and conserving field theoretic technique for describing the far-from-equilibrium quantum dynamics of strongly interacting spin-$1/2$ systems for arbitrary lattices and initial states. Referred to as the Spin-2PI formalism, this method systematically improves upon the mean-field description by including quantum fluctuations by means of an asymptotic $1/N$ expansion, which is controlled in models with intrinsically large lattice coordination number.

We utilized the Spin-2PI technique to study far-from-equilibrium phenomena in spin systems with continuous symmetries. Specifically, we explored the relaxation dynamics of spin spiral states in the 3D Heisenberg model, treating the spiral winding $Q$ as a tuning parameter. Going beyond the trivial mean-field (LO) dynamics by including the NLO correction, we found the spiral states with different windings to relax in remarkably different ways. In particular, spiral states resembling FM and AFM ordered states, corresponding to $Q \sim 0$ and $\pi$ respectively, get trapped for long times in non-thermal states, i.e. ``false vacuums'' whose lifetime diverge as the windings are tuned to $Q=0$ or $\pi$. In contrast, the spiral states far from $Q=0,\pi$ relax rapidly.

We calculated the effective temperature of spin and exchange field fluctuations from the fluctuation-dissipation relation. For $Q \sim \pi/2$ spiral states, all modes reach a single temperature, supporting full thermalization in accordance with the eigenstate thermalization hypothesis. In contrast, the different bosonic modes of prethermalizing spirals settle at different temperatures.

We investigated the dynamical formation of correlations and found that the collective modes predicted to be unstable from a linear response analysis, are self-regularized at rather short time scales, demonstrating the importance of the nonlinear effects and non-perturbative treatments. The growth of out-of-plane fluctuations cause the eventual decay of the prethermal states. The restoration of $SU(2)$ symmetry occurs much later after the decay of magnetization, suggesting a hierarchical relaxation reminiscent of coarsening and aging in classical glassy systems. Our results can be tested readily in ultracold atoms experiments with two-component Mott insulators in 3D optical lattices, such as a 3D extension of the experiments in Refs.~\cite{hild_far--equilibrium_2014,brown_2d_2014}.\\

This work can be extended in several directions.
A straightforward extension is to investigate the relaxation of spiral states in anisotropic models, or in lower dimensions. Another immediately accessible direction is to study spin systems with long-range interactions, as realized for instance with Rydberg atoms, polar molecules, or trapped ions, and their instability toward dynamic crystallization.
The effect of small deviations from the initial spiral state on the quantum evolution could be studied as well. We expect our predictions to carry over to the case of weakly disordered initial states, provided that the deviations from the pristine spiral state remain small by the time the prethermalization plateau is reached. A conservative bound for the allowed degree of disorder can be estimated from the presented linear response analysis, however, a more realistic calculation must take into account self-regulation and slowing down of the unstable modes in the presence of disorder, which is a computationally challenging task.
On related grounds, it is desirable to study the formation of topological defects in quenches to the ordered phase, corresponding to the instantaneous limit of the quantum Kibble-Zurek mechanism~\cite{kibble_topology_1976,zurek_cosmological_1985}. For the 3D Heisenberg model with $SU(2)$ symmetry, topologically stable hedgehogs~\cite{chaikin_principles_2000} are expected to form with universal scaling laws. Other possible research directions include extension to open spin systems, studying the role of NNLO corrections to assess the robustness of the NLO results, and comparison with other systematic expansions such as $1/D$-expansion in $D$-dimensional lattices, and the semi-classical $1/S$-expansion.

\section{Acknowledgements}

We thank S. Gopalakrishnan, B.I. Halperin, and S. Sachdev for useful discussions. We acknowledge support from Harvard-MIT CUA, NSF Grant No. DMR-1308435, AFOSR Quantum Simulation MURI, the ARO-MURI on Atomtronics, ARO MURI Quism program, Humboldt Foundation, the Institute for Quantum Information and Matter, an NSF Physics Frontiers Center with support of the Gordon and Betty Moore Foundation, the Austrian Science Fund (FWF) Project No. J 3361-N20 as well as  Technische Universit\"at M\"unchen - Institute for Advanced Study, funded by the German Excellence Initiative and the European Union FP7 under grant agreement 291763. ED acknowledges support from Dr.~Max R\"ossler, the Walter Haefner Foundation and the ETH Foundation.

\appendix

\section{Summary of the truncated Spin-2PI formalism at the NLO level}\label{sec:supp}
In this Appendix, we provide supplementary material for the Spin-2PI formalism along with a brief account of the numerical methods. The covered material includes the explicit derivation of the approximate dynamical equations from the NLO truncated 2PI effective action and the reconstruction of real-time spin-spin correlators from the Bethe-Salpeter equation.

\subsection{Correlation functions on the Schwinger-Keldysh time contour}\label{sec:CTPstruct}
In the Schwinger-Keldysh formalism, the non-equilibrium dynamics of quantum fields is most elegantly derived from a path-integral defined on the round-trip contour $\CC = \CC^+ \cup \CC^-$:
\begin{equation*}
\eqfigscl{1.4}{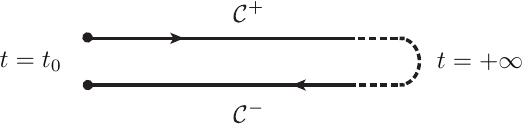}
\end{equation*}
The Majorana operators $\eta$ and real vector boson $\phii$ are replaced by Grassmann and real vector valued variables in the path-integral, along with an anti-periodic and periodic boundary condition at the contour endpoints, respectively. 

The correlation functions defined on the $\CC$ contour can be thought of $2 \times 2$ matrices in the two-dimensional space of the contour branch index. For example, the Majorana $2$-point correlator $\GG$ can be explicitly written as:
\begin{equation}
\GG(t_1, t_2) = \left(
\begin{array}{cc}
\GG^{++}(t_1, t_2) & \GG^{+-}(t_1, t_2)\\
\GG^{-+}(t_1, t_2) & \GG^{--}(t_1, t_2)
\end{array}\right),
\end{equation}
where the times appearing in the matrix are ordinary times. We have dropped the discrete indices for brevity. The off-diagonal matrix elements are identified with the ``lesser'' and ``greater'' explicitly ordered correlators:
\begin{align}\label{eq:Gpmmp}
\GG^{+-}(t_1, t_2) &\equiv \GG^<(t_1, t_2) = +i\big\langle \eta(t_2) \, \eta(t_1) \big\rangle,\nonumber\\
\GG^{-+}(t_1, t_2) &\equiv \GG^>(t_1, t_2) = -i\big\langle \eta(t_1) \, \eta(t_2) \big\rangle.
\end{align}
The diagonal matrix elements are related to each other by the virtue of the unitarity of evolution:
\begin{align}\label{eq:Ggtrless}
&\GG^{++}(t_1, t_2) = +\theta(t_1-t_2)\big[\GG^{>}(t_1, t_2) - \GG^{<}(t_1, t_2)\big],\nonumber\\
&\GG^{--}(t_1, t_2) = -\theta(t_2-t_1)\big[\GG^{>}(t_1, t_2) - \GG^{<}(t_1, t_2)\big],
\end{align}
which are identified with the usual retarded and advanced response functions, $\GG^{++} \equiv \GG^R$ and $\GG^{--} \equiv \GG^A$. While the lesser and greater correlation functions are independent functions for Dirac (complex) fermions, they are related to each other for Majorana fermions by transposition and negation, as it can be seen from Eq.~\eqref{eq:Ggtrless}:
\begin{equation}
\GG^>(1,2) = - \GG^<(2,1).
\end{equation}
In summary, the $2$-point correlator of Majorana fermions on the contour is fully specified by a single real-time correlator, e.g. $\GG^>(1,2)$. It is easily shown that the same decomposition and relations hold for the Majorana self-energy $\Sigma$. The correlator of real bosons $\DD$ and the bosonic self-energy $\Pi$ admit a similar decomposition, except for the absence of the relative minus sign in the definition of $\DD^>$ and $\DD^<$:
\begin{align}\label{eq:Dpmmp}
\DD^{+-}(t_1, t_2) &\equiv \DD^<(t_1, t_2) = -i\big\langle \phii(t_2) \, \phii(t_1) \big\rangle,\nonumber\\
\DD^{-+}(t_1, t_2) &\equiv \DD^>(t_1, t_2) = -i\big\langle \phii(t_1) \, \phii(t_2) \big\rangle,
\end{align}
which imply:
\begin{equation}
\DD^>(1,2) = \DD^<(2,1).
\end{equation}
Similar to Majorana correlators, the $2$-point correlator of real bosons on the contour is fully specified by a single real-time correlator, e.g. $\DD^>$. The same result holds for the bosonic self-energy $\Pi$.

\subsection{Feynman rules for the Spin-2PI formalism}\label{app:feyn}
The conventional Feynman diagram rules are used for interpreting the diagrams appearing throughout this work:
\begin{equation*}
\includegraphics[scale=1.4]{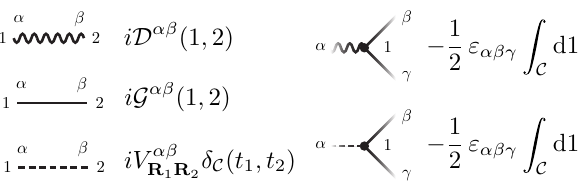}
\end{equation*}
The integer indices refer to the bundle of lattice site and contour time in the diagrams above. Since the Majorana fermion propagators possess no charge flow direction, one may arbitrarily assign a direction to each line. The overall sign of each diagram, however, must be determined at the end by counting the number of fermion permutations. 

The power counting of the large-$N$ extension is performed as follows: (1) each Majorana fermion loop introduces a factor of $N$ resulting from the replica summation, (2) each interaction and boson line introduces a factor of $1/N$.

The vacuum diagrams accompany symmetry factors which must be worked out case by case. The self-energy $\Sigma, \Pi$ and 4-point vertex $\Lambda^{(2)}$ diagrams have an extra factor of $i$ and $i^2$, respectively.

\subsection{Evolution of correlations functions in the Spin-2PI formalism}\label{app:details}

The transition from the path-integral to the 2PI effective action $\GAM[\GG,\DD, \Phii]$ was briefly outlined in the main text and is a straightforward generalization of the results of Cornwall, Jackiw, and Tomboulis~\cite{cornwall_effective_1974}. Within this formalism, the evolution equations follow from a variational principle, reminiscent of Lagrangian dynamics of classical particles, with the quantum correlators playing the role of generalized coordinates. Going back to Eqs.~\eqref{eq:gam} and~\eqref{eq:gam2} and making $\GAM$ stationary with respect to $\GG$, $\DD$, and $\phii$, we obtain:
\begin{subequations}
\begin{align}
\label{eq:SMdys1}
\GG^{-1} &= \GG_0^{-1} - M[\Phii_c] - \Sigma[\GG,\DD],\\
\label{eq:SMdys2}
\DD^{-1} &=  \DD_0^{-1} - \Pi[\GG,\DD],\\
\label{eq:SMdys3}
\phii^\mu_{c,j}(t) &= \frac{1}{2N}\sum_{\sigma=1}^N V^{\mu\nu}_{jk} \epsilon_{\nu\gamma\lambda}\,\GG^{\gamma;\sigma, \lambda;\sigma}_{jj}(t^+,t).
\end{align}
\end{subequations}
With the spin, replica, time, and space indices laid out explicitly, the ``bare'' fermion and boson propagators are written as:
\begin{align}
\GG_0^{-1}(1,2) &=\delta_{\sigma_1\sigma_2}\delta_{\alpha_1\alpha_2}\delta_{j_1 j_2}\,i\partial_{t_1}\delta_\CC(t_1,t_2),\nonumber\\
\DD_0^{-1}(1,2) &= N\, (V^{-1})^{\alpha_1\alpha_2}_{j_1 j_2}\,\delta_\CC(t_1,t_2),
\end{align}
respectively. The contour Dirac $\delta$-function is defined as $\delta_\CC(t_1,t_2) = \pm \delta(t_1 - t_2)$ with the $\pm$ sign corresponding to $t_1, t_2 \in \CC^\pm$, respectively. In Eq.~\eqref{eq:SMdys1}, $M[\Phii_c](1,2) = -i\delta_{\sigma_1\sigma_2}\delta_\CC(t_1,t_2^+)\,\delta_{j_1 j_2}\,\phii^\mu_{c,j_1}(t_1)\,\epss_{\mu\alpha_1\alpha_2}$ the LO interaction effect and describes the coupling of Majorana fermions with the classical spin mean-field $\Phii_c$. According to Eq.~\eqref{eq:SMdys3}, the latter is instantaneously determined by the Majorana tadpole contracted with a bare interaction line. Thus, we find:
\begin{multline}\label{eq:Mdef}
\hspace{-5pt}
M[\Phii_c](1,2) = -i\,\delta_{\sigma_1\sigma_2} \, \delta_\CC(t_1,t_2^+) \, \delta_{j_1 j_2} \, \epss_{\alpha_1\alpha_2\mu} \, V^{\mu\nu}_{j_1 k} \, \epsilon_{\nu\gamma\lambda}\\
\times \frac{1}{2N}\sum_{\sigma=1}^N \GG^{\gamma;\sigma, \lambda;\sigma}_{kk}(t_1^+,t_1) \, = 2 \times \, \eqfigscl{1.2}{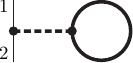}\,\,,
\end{multline}
which resembles the familiar Hartree self-energy that describes the mean-field effects. We emphasize that the Majorana tadpole is identified with the magnetization in our formalism. Note that the $M(1,2) \propto \delta_\CC(t_1,t_2^+)$ is instantaneous and carries no memory effect. We will later show that truncating the approximation at this level and neglecting the self-energies indeed yields the Bloch equation.\\

Going beyond the LO approximation, $\Sigma[\GG,\DD]$ and $\Pi[\GG,\DD]$ describe memory effects associated from the spatiotemporal fluctuations of the exchange field. By definition, these self-energies are obtained from the variations of $\GAM_2[\GG,\DD]$:
\begin{align}
\Sigma[\GG,\DD](1,2) &\equiv 2 \, \frac{\delta \GAM_2[\GG,\DD]}{\delta \GG(1,2)},\nonumber\\
\Pi[\GG,\DD](1,2) &\equiv 2 \, \frac{\delta \GAM_2[\GG,\DD]}{\delta \DD(1,2)}.
\end{align}
We recall that $\GAM_2[\GG,\DD]$ is formally equivalent to the sum of 2PI vacuum diagrams constructed from the interaction vertex $\LL_{\rm int}[\eta,\Phii] = \frac{i}{2} \, \epss_{\alpha\beta\gamma}\,\phii^\alpha_j\,\eta^{\beta;\sigma}_j\eta^{\gamma;\sigma}_j$ and admits a systematic expansion in $1/N$. Here, we truncate the series at the NLO level:
\begin{align}
&\GAM^{\rm NLO}_2[\GG,\DD] = \frac{1}{4}\tr[\DD \Pi_0] = \eqfigscl{0.8}{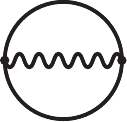}\,,
\end{align} 
where $\Pi^{\mu\nu}_0(1,2) = i\,\epss_{\mu\alpha\beta}\,\epss_{\nu\gamma\lambda}\,\sum_{\sigma=1}^N \GG_{j_1 j_2}^{\alpha;\sigma, \gamma;\sigma}(t_1,t_2)\allowbreak\,\GG_{j_1 j_2}^{\beta;\sigma, \lambda;\sigma}(t_1,t_2)$ is the Majorana bubble. Since $\DD \sim 1/N$ and the factor of $N$ resulting from the replica summation in the Majorana bubble, we find $\GAM^{\rm NLO}_2 \sim \mathcal{O}(1)$. This must be compared to the LO term in $\GAM_{\rm int}$ which is $\mathcal{O}(N)$. The resulting NLO self-energies are given as:
\begin{align}\label{eq:senNLO}
\Sigma^{\rm NLO}(1,2) &= 4 \times \, \eqfigsclbot{1.2}{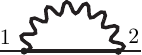}\nonumber\\
&=i\,\epss_{\alpha_1\beta_1\mu} \, \DD^{\mu\nu}_{j_1 j_2}(t_1, t_2) \, \epss_{\nu \alpha_2 \beta_2} \, \GG^{\beta_1\beta_2}_{j_1 j_2}(t_1,t_2),\nonumber\\
\Pi^{\rm NLO}(1,2) & = 2 \times \, \eqfigscl{1.2}{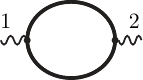} \, = \frac{1}{2}\,\Pi_0(1,2).
\end{align}

Having derived the explicit expressions for the self-energies, we discuss the derivation of evolution equations as the next step. Our starting point are the coupled Dyson's equations given in Eqs.~\eqref{eq:SMdys1} and~\eqref{eq:SMdys2}. Strictly speaking, Dyson's equations are differential identities on the contour functions. They can be cast into a more useful form by acting them from the left and right hand side by $\GG$ and $\DD$, respectively, resulting in a set of contour integro-differential equations:
\allowdisplaybreaks
\begin{widetext}
\begin{subequations}
\begin{align}
\label{eq:KB1}
&\big[i\delta_{\alpha_1\mu}\,\delta_{j_1 k}\,\partial_{t_1} + i\phii_{c,k}^{\nu}(t_1) \, \epss_{\nu\alpha_1\mu}\big] \, \GG^{\mu \alpha_2}_{k j_2}(t_1,t_2) = \delta(1,2) + \int_\CC \dd \tau \, \Sigma^{\alpha_1 \mu}_{j_1 k}(t_1,\tau) \, \GG^{\mu \alpha_2}_{k j_2}(\tau,t_2),\\
\label{eq:KB2}
-&\big[i\delta_{\mu\alpha_2}\,\delta_{k j_2}\partial_{t_2} - i\phii_{c,k}^{\nu}(t_1)\,\epss_{\nu\mu\alpha_2}\big] \, \GG^{\alpha_1 \mu}_{j_1 k}(t_1,t_2) = \delta(1,2) + \int_\CC \dd \tau \, \GG^{\alpha_1 \mu}_{j_1 k}(t_1,\tau) \, \Sigma^{\mu \alpha_2}_{k j_2}(\tau,t_2),
\end{align}
\end{subequations}
\begin{subequations}
\begin{align}
\label{eq:D1}
&\DD^{\alpha_1\beta_1}_{j_1 j_2}(t_1,t_2) = \frac{1}{N}\,V^{\alpha_1\alpha_2}_{j_1 j_2}\,\delta_\CC(t_1, t_2) + \frac{1}{N}\,V^{\alpha_1\mu}_{j_1 k}\,\int_\CC \dd \tau \, \Pi^{\mu\nu}_{kl}(t_1,\tau) \, \DD^{\nu\alpha_2}_{lj_2}(\tau, t_2),\\
\label{eq:D2}
&\DD^{\alpha_1\beta_1}_{j_1 j_2}(t_1,t_2) = \frac{1}{N}\,V^{\alpha_1\alpha_2}_{j_1 j_2}\,\delta_\CC(t_1, t_2) + \frac{1}{N}\,\int_\CC \dd \tau \, \DD^{\alpha_1\mu}_{j_1 k}(t_1,\tau) \, \Pi^{\mu\nu}_{kl}(\tau, t_2)\, V^{\nu\alpha_2}_{l j_2}.
\end{align}
\end{subequations}
\end{widetext}
We have defined shorthand $\delta(1,2) \equiv \delta_{j_1 j_2} \, \delta_{\alpha_1\alpha_2} \, \delta_\CC(t_1,t_2)$ and the contour integral $\int_\CC \dd t \, \mathcal{A}(t)$ is interpreted as $\int_{t_0}^\infty \dd t \, \mathcal{A}(t \in \CC^+)- \int_{t_0}^\infty \dd t \, \mathcal{A}(t \in \CC^-)$. Eq.~\eqref{eq:KB1} and its adjoint Eq.~\eqref{eq:KB2} are referred to as Kadanoff-Baym (KB) equations. The convolution integrals of self-energies and correlators manifestly show memory effects, which is a shared feature of beyond mean-field approximations.

The spatial structure of Eqs.~\eqref{eq:KB1}-\eqref{eq:D2} can be  simplified by noting that physical initial states imply initial correlations between pairs of Majorana operators on the same site, i.e. $\GG_{j_1 j_2}^{\alpha_1\alpha_2}(t_0, t_0) \propto \delta_{j_1 j_2}$. Crucially, this property extends to all times in the KB dynamics, independent of the order of truncation in $1/N$. To see this, one first establishes that the assumption $\GG_{j_1 j_2}(t_1,t_2) \propto \delta_{j_1 j_2}$ for $ t_0 \leq t_1, t_2 \leq T$ implies $\Sigma_{j_1 j_2}(t_1,t_2) \propto \delta_{j_1 j_2}$ in the same domain. The causal structure of Eqs.~\eqref{eq:KB1}-\eqref{eq:KB2} subsequently extends this property to an infinitesimally larger domains, and eventually to all times by induction. Therefore, we can always make the following simplifying substitution in the KB equation:
\begin{align}\label{eq:Gstruct1}
\GG_{j_1 j_2}^{\alpha_1 \alpha_2}(t_1, t_2) &\rightarrow \delta_{j_1 j_2} \, \GG_{j_1 j_1}^{\alpha_1\alpha_2}(t_1, t_2),\nonumber\\
\Sigma_{j_1 j_2}^{\alpha_1 \alpha_2}(t_1, t_2) &\rightarrow \delta_{j_1 j_2} \, \Sigma_{j_1 j_1}^{\alpha_1\alpha_2}(t_1, t_2).
\end{align}

\noindent{\em The LO approximation:} The KB equations reduce to the mean-field Bloch equation upon truncation at the LO level which amounts to neglecting fluctuation self-energy corrections $\Sigma \rightarrow 0$. In this limit, the KB equations imply:
\begin{align}
&i\partial_{t_1} \GG^{\alpha_1\alpha_2, >}_{jj}(t_1, t_2) + i\phii^\nu_{c,j}(t_1)\,\epss_{\nu\alpha_1\mu}\,\GG^{\mu\alpha_2}_{jj}(t_1,t_2) = 0,\nonumber\\
-&i\partial_{t_2} \GG^{\alpha_1\alpha_2, >}_{jj}(t_1, t_2) + i\phii^\nu_{c,j}(t_2)\,\epss_{\nu\mu\alpha_2}\,\GG^{\alpha_1\mu}_{jj}(t_1,t_2) = 0.
\end{align}
Subtracting the equations from one another, setting $t_2 = t_1 = t$ and using \eq{eq:SG}, we finally obtain:
\begin{equation}
\partial_t \langle \Shat_j(t) \rangle = \Phii_{c,j} \times \langle \Shat_j(t)\rangle,
\end{equation}
which is the Bloch equation as anticipated.\\

\noindent{\em The NLO approximation:} Including self-energy corrections, the time convolutions appearing in the KB equations prohibit us from arriving at a closed equation for the equal-time Green's functions and we inevitably need to solve for the complete unequal time Green's function. For concreteness, we consider the case of spin spirals hereafter. The spatial structure of the KB equations can be significantly simplified by applying the unwinding unitary transformation, either directly on Eqs.~\eqref{eq:KB1}-\eqref{eq:D2} or on the spin Hamiltonian. Either way, the initial spiral state transforms into an uncorrelated $x$-polarized FM state $|\tilde{\Psi}_0 \rangle \equiv \bigotimes_j |\rightarrow\rangle_j$ at the expense of an anisotropic interaction (see Eq.~\ref{eq:Hspiral}). The $2$-point correlator of Majorana fermions at $t=t_0$ is easily found as:
\begin{equation}\label{eq:G0}
\GG^{\alpha_1\alpha_2,>}_{j_1 j_2}(t_0,t_0) = \delta_{j_1 j_2}\left(\begin{tabular}{cccc}
$-i/2$ & 0 & 0 & $-i/2$\\
0 & $-i/2$ & $1/2$ & 0\\
0 & $-1/2$ & $-i/2$ & 0\\
$-i/2$ & 0 & 0 & $-i/2$
\end{tabular}\right).
\end{equation}
The exchange field correlator at $t=t_0$ is not an independent degree of freedom and is determined by $\GG(t_0, t_0)$, see \eq{eq:senNLO}. For translationally invariant initial states as such, $\GG$ and $\Sigma$ further become independent of the lattice site. Furthermore, $\DD_{j_1 j_2}$ depends only on the distance and at the NLO level, $\Pi_{j_1 j_2}$ is local as well. The simplified structure of the correlators and self-energies is summarized as follows:
\begin{align*}
\GG(1,2) &\rightarrow \delta_{\RR_1\RR_2} \, G^{\alpha\beta}(t_1,t_2),\\
\Sigma(1,2) &\rightarrow \delta_{\RR_1\RR_2} \, \Sigma^{\alpha\beta}(t_1,t_2),\\
\DD(1,2) &\rightarrow \DD^{\alpha\beta}_{\RR_1 - \RR_2}(t_1,t_2) \xrightarrow{\text{F.T.}} \DD^{\alpha\beta}_{\kk}(t_1,t_2),\\
\Pi(1,2) &\rightarrow \delta_{\rr_1\rr_2} \, \Pi^{\alpha\beta}(t_1,t_2).
\end{align*}

The KB equations can be written explicitly in terms of $\GG^>$ and $\DD^>$ using the Langreth rules~\cite{rammer_quantum_1986}. We quote the final result, setting $N=1$:
\begin{widetext}
\begin{subequations}
\begin{align}
\label{eq:KB3}
i\partial_{t_1} \GG^{\alpha_1\alpha_2}(t_1, t_2) + i\phii_c^{\alpha_1\mu}(t_1)\,\GG^{\mu\alpha_2,>}(t_1,t_2) =& \int_{t_0}^{t_1}\dd \tau\,\left[\Sigma^{\alpha_1\mu,>}(t_1,\tau) + \Sigma^{\mu\alpha_1,>}(\tau,t)\right]\,\GG^{\mu\alpha_2,>}(\tau,t_2)\nonumber\\
&-\int_{t_0}^{t_2}\,\dd \tau\,\Sigma^{\alpha_1\mu,>}(t_1,\tau)\left[\GG^{\mu\alpha_2,>}(\tau,t_2)+\GG^{\alpha_2\mu,>}(t_2,\tau)\right],\\
\label{eq:KB4}
-i\partial_{t_2} \GG^{\alpha_1\alpha_2}(t_1, t_2) + i\GG^{\alpha_1\mu,>}(t_1,t_2) \, \phii_c^{\mu\alpha_2}(t_2) =& \int_{t_0}^{t_1}\dd \tau\,\left[\GG^{\alpha_1\mu,>}(t_1,\tau) + \GG^{\mu\alpha_1,>}(\tau,t)\right]\,\Sigma^{\mu\alpha_2,>}(\tau,t_2)\nonumber\\
&-\int_{t_0}^{t_2}\,\dd \tau\,\GG^{\alpha_1\mu,>}(t_1,\tau)\left[\Sigma^{\mu\alpha_2,>}(\tau,t_2)+\Sigma^{\alpha_2\mu,>}(t_2,\tau)\right],
\end{align}
\end{subequations}
\begin{subequations}
\begin{align}
\label{eq:D3}
\DD_\kk^{\alpha_1\alpha_2,>}(t_1,t_2) = V_\kk^{\alpha_1\mu}\,\Pi^{\mu\nu,>}(t_1,t_2)\,V_\kk^{\nu\alpha_2} &+ V_\kk^{\alpha_1\mu}\int_{t_0}^{t_1}\dd \tau \left[\Pi^{\mu\nu,>}(t_1,\tau)-\Pi^{\nu\mu,>}(\tau,t_1)\right]\DD_\kk^{\nu\alpha_2,>}(\tau,t_2)\nonumber\\
&-V_\kk^{\alpha_1\mu}\int_{t_0}^{t_2}\,\dd \tau\,\Pi^{\mu\nu,>}(t_1,\tau)\left[\DD_\kk^{\nu\alpha_2,>}(\tau,t_2) - \DD_\kk^{\alpha_2\nu,>}(t_2,\tau)\right],\\
\label{eq:D4}
\DD_\kk^{\alpha_1\alpha_2,>}(t_1,t_2) = V_\kk^{\alpha_1\mu}\,\Pi^{\mu\nu,>}(t_1,t_2)\,V_\kk^{\nu\alpha_2} &+ \int_{t_0}^{t_1}\dd \tau \left[\DD_\kk^{\alpha_1\mu,>}(t_1,\tau)-\DD_\kk^{\mu\alpha_1,>}(\tau,t_1)\right]\Pi^{\mu\nu,>}(\tau,t_2)\,V_\kk^{\nu\alpha_2}\nonumber\\
&-\int_{t_0}^{t_2}\,\dd \tau\,\DD_\kk^{\alpha_1\mu,>}(t_1,\tau)\left[\Pi^{\mu\nu,>}(\tau,t_2) - \Pi^{\nu\mu,>}(t_2,\tau)\right] V_\kk^{\nu\alpha_2}.
\end{align}
\end{subequations}
\end{widetext}
The self-energies $\Sigma^>$ and $\Pi^>$ are read from Eq.~\eqref{eq:senNLO}. The last explicit equations are suitable for devising a numerical forward propagation scheme. Starting from $\GG_0(t_0, t_0)$, we calculate $\Sigma(t_0,t_0)$ and $\Pi(t_0, t_0)$ from Eq.~\eqref{eq:senNLO}, and $\DD(t_0,t_0)$ from Eq.~\eqref{eq:D3}. The casual structure Eqs.~\eqref{eq:KB3}-\eqref{eq:D4} allows us to propagate $\{\GG,\Sigma,\DD,\Pi\}$ in $(t_1, t_2)$ in discrete time steps of size $\Delta t$. This is achieved using a robust predictor-corrector method with guaranteed accuracy to $\mathcal{O}(\Delta t^3)$.\\

\noindent{\it Calculating spin-spin correlators in the Spin-2PI~formalism:} In the framework of 2PI effective actions, higher order correlators  are ``reconstructed'' from the history of $2$-point correlators. Here, we are interested in the connected spin-spin correlator:
\begin{multline}
i\chi^{\alpha_1\alpha_2}_{j_1 j_2}(t_1,t_2) = \left\langle T_\CC \big[ \shat^{\alpha_1}_{j_1}(t_1) \shat^{\alpha_2}_{j_2}(t_2) \big] \right\rangle \\- \left\langle \shat^{\alpha_1}_{j_1}(t_1) \right\rangle \, \left\langle \shat^{\alpha_2}_{j_2}(t_2) \right\rangle,
\end{multline}
where the spin operators are shorthand notations for Eq.~\eqref{eq:spinrep}. The spin-spin correlator is found from the Majorana $L$-function, defined as:
\begin{equation}
L(1\bar{1};2\bar{2}) \equiv \left\langle T_\CC \left[\eta(1) \, \eta(\bar{1}) \, \eta(2) \, \eta(\bar{2}) \right] \right\rangle - i\GG(1,\bar{1})\,i\GG(2,\bar{2}),
\end{equation}
by contracting $\epss$-symbols with its left and right pair of fermion lines:
\begin{equation}\label{eq:SMchiL}
\chi^{\alpha_1\alpha_2}_{j_1 j_2}(t_1, t_2) =  \frac{i}{4}\, \epss_{\alpha_1\beta_1\gamma_1} \, L^{\beta_1\gamma_1;\beta_2\gamma_2}_{j_1 j_1; j_2 j_2}(t_1^+,t_1^{\phantom{+}}; t_2^+, t_2^{\phantom{+}})\, \epss_{\alpha_2\beta_2\gamma_2}.
\end{equation}
The $L$-function in turn satisfies a non-equilibrium Bethe-Salpeter equation on the $\CC$ contour:
\begin{multline}\label{eq:SMBSE}
L(1\bar{1};2\bar{2}) = \Pi_2(1\bar{1};2\bar{2})\\ + \int_\CC \dd 3 \, \dd \bar{3} \, \dd 4 \, \dd\bar{4}\,\Pi_2(1\bar{1};3\bar{3})\,\Lambda^{(2)}(3\bar{3};4\bar{4})\,L(4\bar{4};2\bar{2}),
\end{multline}
where $\Pi_2(1\bar{1};2\bar{2}) = \GG(12) \, \GG(\bar{2}\bar{1}) - \GG(1\bar{2}) \, \GG(\bar{1}2)$ and the 2PI irreducible vertex $\Lambda^{(2)}(3\bar{3};4\bar{4}) = \delta^2 \GAM_\mathrm{int}[\GG] / \delta \GG(3\bar{3}) \delta \GG(4\bar{4})$; here, $\GAM_\mathrm{int}[\GG]$ is given in Eq.~\eqref{eq:gam2} with $\Phii_c$ and $\DD$ substituted in terms of $\GG$ from the stationarity condition Eqs.~\eqref{eq:SMdys2} and~\eqref{eq:SMdys3}.

The NLO effective action yields three contributions to $\Lambda^{(2)}$:
\begin{equation}
\Lambda^{(2)} = \frac{1}{4} \times \underbrace{\,\eqfigscl{0.75}{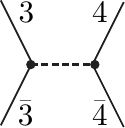}\,}_{\Lambda^{(2)}_\mathrm{RPA}} - \frac{1}{2} \times \underbrace{\,\eqfigscl{0.75}{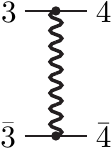} \,}_{\Lambda^{(2)}_\mathrm{MT}} + 8 \times \underbrace{\,\eqfigscl{0.75}{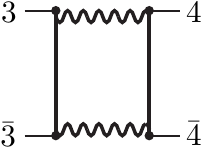}\,}_{\Lambda^{(2)}_\mathrm{AL}} + \,\,\,\eqfigscl{0.85}{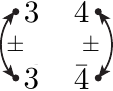}\,.
\end{equation}
The last symbol stands for the three permutations of the first three diagrams obtained by $(3 \leftrightarrow \bar{3})$, $(4 \leftrightarrow \bar{4})$, and $(3 \leftrightarrow \bar{3}, 4 \leftrightarrow \bar{4})$ with signs $-$, $-$ and $+$, respectively. These vertex corrections are structurally similar to the RPA, Maki-Thompson (MT), and Aslamazov-Larkin (AL) vertex corrections accounting for superconducting fluctuations in metals~\cite{varlamov1997fluctuation}. Explicitly, these vertex parts are given as:
\begin{subequations}
\begin{align}
\Lambda^{(2)}_{\rm RPA}(3\bar{3}; 4\bar{4}) =&\, \frac{1}{2}\,\epss_{\alpha_3 \alpha_{\bar{3}}\mu}\,iV_{j_3 j_4}^{\mu\nu}\,\frac{1}{2}\, \epss_{\nu \alpha_4 \alpha_{\bar{4}}}\,\delta_{j_3 j_{\bar{3}}}\,\delta_{j_4 j_{\bar{4}}}\nonumber\\
&\times \delta_\CC(t_3,t_{\bar{3}})\,\delta_\CC(t_4,t_{\bar{4}})\,\,\delta_\CC(t_3,t_4),\\
\Lambda^{(2)}_{\rm MT}(3\bar{3}; 4\bar{4}) =&\, \frac{1}{2}\,\epss_{\alpha_3 \alpha_4\mu}\,i\DD_{j_3 j_{\bar{3}}}^{\mu\nu}(t_3,t_{\bar{3}})\, \frac{1}{2}\,\epss_{\nu \alpha_{\bar{3}} \alpha_{\bar{4}}}\,\delta_{j_3 j_4}\,\delta_{j_{\bar{3}} j_{\bar{4}}}\nonumber\\
&\times \delta_\CC(t_3,t_4)\,\delta_\CC(t_{\bar{3}},t_{\bar{4}}),\\
\Lambda^{(2)}_{\rm AL}(3\bar{3}; 4\bar{4}) =&\, i\GG_{j_3 j_{\bar{3}}}^{\beta_3\beta_{\bar{3}}}(t_3,t_{\bar{3}})\,i\GG_{j_{4}j_{\bar{4}}}^{\beta_{4}\beta_{\bar{4}}}(t_{4},t_{\bar{4}})\,\nonumber\\
&\times \frac{1}{2}\,\epss_{\alpha_3\beta_3\mu}\,i\DD^{\mu\nu}_{j_3 j_4}(t_3, t_4) \, \frac{1}{2}\,\epss_{\nu\alpha_4\beta_4}\nonumber\\
&\times \frac{1}{2}\,\epss_{\alpha_{\bar{3}}\beta_{\bar{3}}\bar{\mu}}\,i\DD^{\bar{\mu}\bar{\nu}}_{j_{\bar{3}} j_{\bar{4}}}(t_{\bar{3}},t_ {\bar{4}}) \, \frac{1}{2}\,\epss_{\bar{\nu}\alpha_{\bar{4}}\beta_{\bar{4}}}
\end{align}
\end{subequations}
It is easily noticed that $\Lambda^{(2)}_{\rm RPA} \sim \mathcal{O}(1)$, $\Lambda^{(2)}_{\rm MT} \sim \mathcal{O}(1/N)$, and $\Lambda^{(2)}_{\rm AL} \sim \mathcal{O}(1/N^2)$. Therefore, we may drop the latter if accuracy at the NLO order is desired.

For translation invariant states, it can be shown that $L(1\bar{1};2\bar{2}) \sim L^{\alpha_1\alpha_{\bar{1}};\alpha_2\alpha_{\bar{2}}}_{\RR_1 - \RR_2}(t_1 t_{\bar{1}};t_2 t_{\bar{2}})\,\delta_{\RR_1 \RR_{\bar{1}}}\,\delta_{\RR_2 \RR_{\bar{2}}}$. Taking a Fourier transform in $\RR_1 - \RR_2$ yields  decoupled integral equations for each momentum transfer $\qq$. The temporal structure of the BSE remains formidable. Performing the contour integral and discrete summations over $3,\bar{3}$ variables in Eq.~\eqref{eq:SMBSE} and contracting the right legs according to Eq.~\eqref{eq:SMchiL}, we reach to an integral equation for the $3$-time object $\Gamma_\qq^{\alpha_1\alpha_{\bar{1}};\mu}(t_1,t_{\bar{1}};t_2) \equiv (-i/2)\,L^{\alpha_1\alpha_{\bar{1}};\alpha_2\alpha_{\bar{2}}}_\qq(t_1, t_{\bar{1}}; t_2^+, t_2)\,\epss_{\mu\alpha_2\alpha_{\bar{2}}}$ in two contour times $t_1, t_{\bar{1}}$ (with fixed external time $t_2$).

The first step in solving the BS equation is to recast it in terms of functions of ordinary times. In comparison to the KB equation, this step is significantly more involved here due to the complex real-time structure of $3$-time and $4$-time CTP functions and multiple contour integrals. We leave the cumbersome details for a separate publication and solely outline the procedure here. We showed earlier in Sec.~\ref{sec:CTPstruct} that the 4 real-time matrix elements of $2$-time functions such as $\GG$ and $\DD$ can be fully specified using a single real-time function, e.g. $\GG^>$. A similar analysis of $\Gamma_\qq^{\alpha_1\alpha_{\bar{1}};\mu}(t_1,t_{\bar{1}};t_2)$, taking into account symmetries and unitarity of evolution, reveals that the 8 real-time components of $3$-time function as such can be fully specified by 3 independent functions. Accordingly, the contour BS equation can be explicitly written as 3 coupled two-dimensional integral equations in ordinary times; the latter is numerically solved by 
discretizing the integrals using approximate quadratures and solving the resulting linear system.

%

\end{document}